\documentclass[aps,prd, twocolumn, nofootinbib, superscriptaddress,amsmath,amssymb,floatfix]{revtex4-2}

\usepackage[dvipdfmx]{graphicx}
\usepackage{dcolumn}
\usepackage{bm}
\usepackage{braket}
\usepackage{breqn}
\usepackage{color}
\usepackage{mathtools} 
\usepackage{hyperref}
\usepackage{comment}
\usepackage{xfp}
\usepackage{siunitx}
\usepackage{subcaption}  
\usepackage{ragged2e}
\newcommand{\mr}[1]{{\mathrm{#1}}}

\newcommand{\pa}{\partial}
\newcommand{\me}{e}
\newcommand{\ec}{e}

\newcommand{\Be}{\begin{equation}}
\newcommand{\Ee}{\end{equation}}

\newcommand{\st}{{}_s }
\newcommand{\resnum}{2612 }

\newcommand{\pE}{126}
\newcommand{\ecE}{0.2}
\newcommand{\xE}{0.7}
\newcommand{\pDM}{25}
\newcommand{\ecDM}{0.2}
\newcommand{\xDM}{0.7}
\newcommand{\aE}{\num{\fpeval{\pE/(1-\ecE^2)}}}
\newcommand{\pEf}{72}
\newcommand{\ecEf}{0.08}
\newcommand{\deltaa}{0.3}

\newcommand{\aEf}{\num{\fpeval{\pEf/(1-\ecEf^2)}}}
\newcommand{\deltaamax}{0.8}

\newcommand{\resscale}{1033}
\newcommand{\tauratio}{\num{\fpeval{100*\resscale/{\resnum}}}}

\begin{document}

\title{Resonant DM scattering in the galactic center under the influence of EMRI}

\author{Takafumi Kakehi} \email{takafumi.kakehi@yukawa.kyoto-u.ac.jp}
 \affiliation{
Yukawa Institute for Theoretical Physics$,$ Kyoto University  
}
\affiliation{Center for Gravitational Physics and Quantum Information$,$ Yukawa Institute for Theoretical Physics$,$ Kyoto University$,$ Kyoto 606-8502$,$ Japan}
\author{Hidetoshi Omiya}
\email{omiya@tap.scphys.kyoto-u.ac.jp}
\affiliation{Department of Physics$,$ Kyoto University$,$ Kyoto 606-8502$,$ Japan}
\author{Takuya Takahashi}\email{takuya.takahashi@resceu.s.u-tokyo.ac.jp}
\affiliation{Research Center for the Early Universe (RESCEU)$,$ Graduate School of Science$,$ The University of Tokyo$,$ Tokyo 113-0033$,$ Japan}
\author{Takahiro Tanaka}
\email{t.tanaka@tap.scphys.kyoto-u.ac.jp}
\affiliation{Department of Physics$,$ Kyoto University$,$ Kyoto 606-8502$,$ Japan}
\affiliation{Center for Gravitational Physics and Quantum Information$,$ Yukawa Institute for Theoretical Physics$,$ Kyoto University$,$ Kyoto 606-8502$,$ Japan}

\date{\today}

\begin{abstract}
Dark matter (DM) search is one of the greatest challenges in physics. If DM consists of particles, it may form a spike around supermassive black holes (BH) prevalent in galaxy centers. This spike could be potentially observed by altering the orbits of Extreme Mass Ratio Inspirals (EMRIs), one of LISA’s main targets. Meanwhile, the effects of EMRI on the DM spike have also been explored. In this study, we focus on the tidal resonances between DM particles and EMRI secondary. As the EMRI orbit evolves via gravitational wave backreaction, each DM particle experiences a significant number of resonances. Although the effect of each individual resonance is small, their cumulative impact might significantly alter the DM particle’s orbit. To examine this possibility, we explicitly derive the interaction Hamiltonian for tidal resonances and conducted numerical calculations.
\end{abstract}

\maketitle

\section{Introduction}
Dark matter (DM) search is one of the current biggest science goals. 
If the DM interaction with the standard model particles is sufficiently weak, we can seek the DM property only through gravity. 
DM may have a high concentration around supermassive black holes (SMBHs), which are ubiquitous at the center of galaxies~\cite{Genzel:2010zy, Iocco:2015xga}.
Particularly high concentration is called DM “spike”~\cite{Gondolo:1999ef, Gnedin:2003rj, Kim:2022mdj}.
In that case extreme mass ratio inspirals (EMRIs)~\cite{Berry:2019wgg, Babak:2017tow} or intermediate mass ratio inspirals (IMRIs)~\cite{Amaro-Seoane:2018gbb, Brown:2006pj}, which are listed in the main targets of space-based gravitational wave (GW) observatories such as LISA~\cite{Colpi:2024xhw, amaroseoane2017laser}, Taiji~\cite{Hu:2017mde}, TianQin~\cite{TianQin:2015yph} and DECIGO~\cite{Kawamura:2011zz}, might be used as a probe of the DM distribution at the galactic center~\cite{Eda:2014kra, Kavanagh:2020cfn}, although EMRI events are usually expected to be a clean system that brings us a detailed map of the spacetime geometry around the central black hole (BH)~\cite{Amaro-Seoane:2012lgq}. 
If DM is densely distributed around the central black hole, the orbits of EMRI will be altered from the vacuum case.
This change is more pronounced if there is a mass concentration, a so-called DM spike, around the central black hole. A DM spike will cause a verifiable shift in the phase of the GWs of the EMRI compared to the vacuum case.
Hence, it is possible to extract information of DM from EMRI observations.

While the DM spike affects the orbital evolution of EMRIs, the DM distribution could also change due to the backreaction from EMRIs. This backreaction, called halo feedback, has been investigated in several previous studies~\cite{Mukherjee:2023lzn, Kavanagh:2020cfn}. 
The effects due to dynamical friction and accretion have been mainly investigated in previous studies~\cite{Becker:2024ibd}. 
It was found that when the mass ratio is close to unity, the DM spike is dissipated by the halo feedback effects~\cite{Kavanagh:2020cfn}. 
However, as far as we know, the extreme mass ratio case has not been extensively investigated.
In this study, we study the behavior of DM particles in the strong gravity region under the influence of EMRI, which has not been explored in detail in previous studies.

Here, we focus on the evolution of the distribution of DM particles in the very vicinity of the galactic center. Without any perturbation from the Kerr background of the central BH, DM particles maintain three constants of motion, and hence, no secular evolution of the distribution occurs. 
However, perturbations must be caused by 
inspiralling satellites for the EMRI/IMRI case as well as by the DM particles themselves. Under these perturbations, the distribution of DM particles may evolve in time. 

The main effect of the EMRI secondary on the DM particles is the direct gravitational two-body scattering. As we shall see below, the two-body 
scattering process is efficient at a large distance, where the secondary object spends a long time because of a weak radiation reaction.

Small perturbations caused by the EMRI secondary at a large distance will not give systematic drift to 
the DM motion, except for the resonant situation~\cite{Bonga:2019ycj, Silva:2022blb}. 
At resonance, some linear combination of orbital frequencies of the EMRI secondary and the DM particle with a set of simple integer coefficients vanishes. 
Then, the perturbations cause the net drift of orbital parameters. Even if the effect of each resonance is tiny, there are many resonances and the cumulative effect is expected to be significant.

In this paper, we analytically and numerically investigate the effects of tidal resonances on DM particles. We will find that DM particles in the LISA band experience a large number of resonances. We will also derive equations that evaluate the effects of the resonances. The cumulative effect of the resonances experienced by DM particles is then investigated numerically for several parameters. We should note that for a technical reason our analysis is restricted to the case the maximum radius of DM particles does not exceed the minimum radius of the EMRI orbit. 

The structure of this paper is as follows. In Sec.~\ref{sec:setup}, we formulate the problem of DM scattering by an EMRI secondary as a restricted three-body problem and explain why tidal resonances are important. 
In Sec.~\ref{sec:derivation}, we derive the interaction Hamiltonian between DM particles and the EMRI secondary. 
In Sec.~\ref{sec:result}, we numerically evaluate the cumulative effects of resonances for typical orbits. In Sec.~\ref{sec:conclusion}, we present the conclusion and future directions.

Throughout this paper, we use a system of units in which the gravitational constant $G$ and the speed of light $c$ are unity, i.e., $G=c=1$.

\section{setup: orbital evolution of EMRI and resonance }
\label{sec:setup}
In this section, we summarize the assumptions on the motion of DM particles. First, we evaluate the direct scattering between DM particles and the EMRI secondary, and show that the resonance effects become important in the vicinity of the central BH in subsection A. 
For simplicity, 
we neglect the DM self-interaction including the self-gravity. 
Namely, we treat the problem as a restricted 3-body problem.
After we review the geodesic motion of particles in Kerr spacetime in subsection B, we formulate the tidal resonance effect of the EMRI secondary on DM particles in subsection C. 

\subsection{Influence of EMRIs in the far region: Direct scattering}
Before studying the resonance effects, we consider the direct scattering. Direct gravitational scattering is modeled by the relaxation processes due to the two-body scattering~\cite{binney2011galactic}.

We denote 
the mass of the central BH and that of the EMRI secondary by 
$M$ and $m$, respectively. 
The orbital velocity, radius, and frequency in the Newtonian circular orbit approximation are given by $v$, $R$, and $\Omega$, respectively.  
They are related to each other as
\begin{align}
  v^2=(\Omega R)^2=\frac{M}{R}\,.
\end{align}
When we consider the direct (=large angle) gravitational scattering of a DM particle by the satellite, the cross section would be roughly estimated by $\sigma=\pi r^2$ with the scattering radius $r$ roughly estimated by 
\begin{align}
  \frac{m}{r}\approx v^2\,.
\end{align}
Thus, we have 
\begin{align}
   \sigma\approx \pi \frac{m^2}{v^4}\,, 
\end{align}
and the volume $V$ that this cross section sweeps in a unit time is roughly estimated by 
\begin{align}
  \frac{dV}{dt}\approx \sigma v\approx \pi \frac{m^2}{v^3}\,.
\end{align}

On the other hand, the orbital radius shrinks 
by the radiation reaction due to GW emission. 
Combining the energy loss rate~\cite{Peters:1963ux}
\begin{align}
    \frac{dE}{dt}\approx -\frac{32 m^2}{5M^2} v^{10}\,,
\end{align}
and the expression for the binding energy $E\approx -M m/2R$, we find 
\begin{align}
 \frac{dR}{dt}=-\frac{64m}{5M}v^6\,.
\end{align}
Hence, the average number of sweeping the volume by the cross section mentioned above is evaluated as 
\begin{align}
 \frac{dV/dt}{4\pi R^2 |dR/dt|}
  = \frac{5 m}{256 Mv^5}\,.
\end{align}

This number exceeds unity for 
\begin{align}
  R> R_{\mathrm{cri}} = 1200 M \left(\frac{m/M}{10^{-6}}\right)^{-2/5}\,.
\end{align}
We find that the direct gravitational scattering is inefficient for the DM particles distributing in the vicinity of the central black hole, which would be relevant to the EMRI motion in the frequency range observable by GWs. 
\begin{figure}
    \centering
    \includegraphics[width=\linewidth]{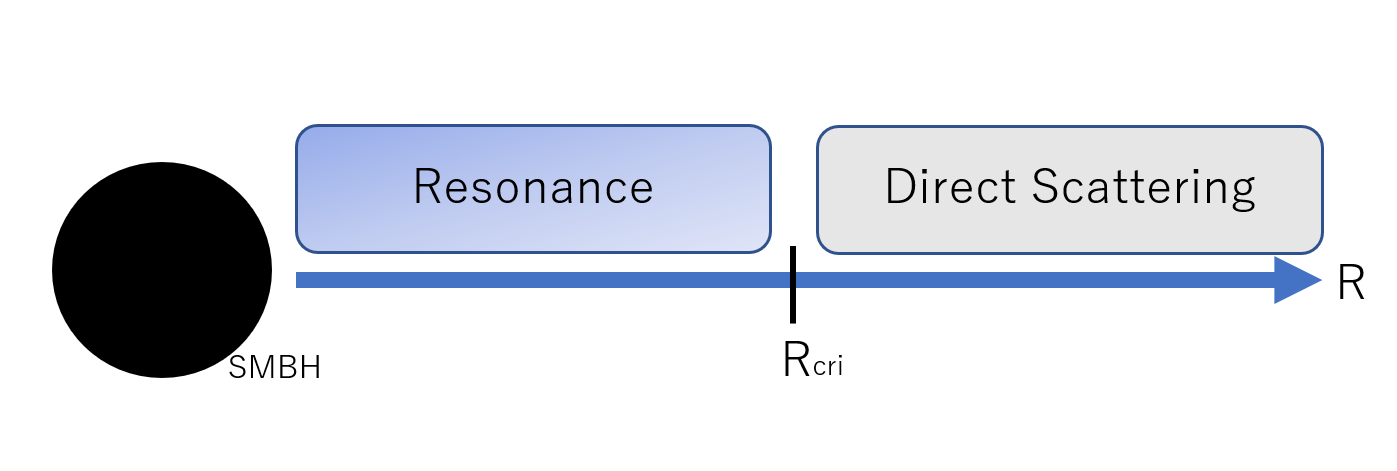}
    \caption{Schematic diagram showing the dominant ranges of direct scattering and resonance, respectively. In the region far from the BH, the direct scattering is dominant, while the effect of resonance becomes important near the BH.}
    \label{fig:directl}
\end{figure}
The effect becomes weaker in the vicinity of the central BH simply because of the shorter orbital evolution timescale relative to the two-body scattering timescale.

From this result, the direct scattering no longer works efficiently within the observational frequency range of GWs when considering the extreme mass ratio regime such as $\eta:=\frac{m}{M}\lesssim 10^{-6}$.
Thus, we consider the resonant effects in the strong gravity region, which is calculated by considering a small perturbation in the Kerr spacetime.

\subsection{Particle motion in Kerr metric}
Before considering the interaction between an EMRI secondary and DM particles, we review the motion of particles in Kerr spacetime~\cite{Gupta:2022jdt}.
We use the metric $g^{(0)}_{\mu\nu}$ expressed in Boyer-Lindquist coordinates $(t,r,\theta,\phi)$,
\begin{align}
&g^{(0)}_{\mu\nu}dx^\mu dx^\nu= -\left(1-\frac{2M r}{\Sigma}\right)dt^2-\frac{4Mar \sin^2\!\theta}{\Sigma}dt\, d\phi \cr &+\frac{\Sigma}{\Delta}dr^2
  +\Sigma\, d\theta^2+\left(r^2+a^2+\frac{2Ma^2 r}{\Sigma}\sin^2\!\theta \right)\sin^2\! \theta\, d\phi^2\,,
\end{align}
with 
\begin{align}
\Sigma= r^2+a^2 \cos^2\!\theta,\qquad \Delta= r^2-2Mr+a^2\,,
\end{align}
and $a$ is the BH spin parameter.

It is known that Kerr geodesics are integrable and there are four constants of motion $(\mu^2, E, L_z, Q)$~\cite{PhysRev.174.1559,Hinderer:2008dm}, which are expressed by
\begin{align}
 &\mu^2=-g^{\mu\nu}p_\mu p_\nu\,,\\ 
 &E =- p_t\,, \\
 &L_z =  p_\phi\,, \\
 &Q =  p_\theta^2 +a^2\cos^2\! \theta (1-p_t^2)+\cot^2\!\theta p_\phi^2\,,
\end{align}
where $p_\nu$ is the conjugate momentum of the particle related to the four-velocity of the particle $u^\nu$ as $p_\nu=\mu u_\nu$. 
 In the following formal discussion, we use a set of action-angle variables~\cite{Schmidt:2002qk}, $(q^\alpha, j_\alpha)$, instead of the pair of the Boyer-Lindquist coordinates and the four momentum $(x^\mu, p_\mu)$. The relation between these two sets of variables is given by the generating function $W(x^\mu,j_\alpha)$, which is defined by
\begin{align}
W(x^\mu,j_\alpha)=&-j_t t+j_\phi \phi \pm \int^r dr' \frac{\sqrt{V_{r}(r^\prime)}}{\Delta}&\cr 
&\qquad \pm \int ^\theta d\theta' \sqrt{V_{\theta}(\theta^\prime)}\,,&
\label{eq:Wxj}
\end{align}
where $V_r$ and $V_\theta$ are given by 
\begin{align}
  V_r (r,E,L_z,Q,\mu^2)=&\left(E(r^2+a^2)-a L_z\right)^2&\cr
   &-\Delta \left[r^2\mu^2+(L_z-a E)^2+Q\right]\,,&\cr
  V_\theta(\theta,E,L_z,Q,\mu^2)=&Q-\cos^2\!\theta\left[a^2(\mu^2-E^2)+\frac{L_z^2}{\sin^2\!\theta}\right]\,.&
  \cr &
\end{align}
By using \eqref{eq:Wxj}, we can express the relation of these variables as
\begin{align}
p_\mu=\frac{\partial W}{\partial x^\mu} (x^\nu,j_\alpha)\,,\\
q^\alpha=\frac{\partial W}{\partial j_\alpha}(x^\nu,j_\alpha)\,.
\end{align}
The expression for $W(x^\mu,j_\alpha)$ contains $(\mu^2,E,L_z,Q)$, which are to be understood as functions of $j_\alpha$ determined by the relations, 
\begin{align}
 & j_t=-E\,,\quad j_\phi=L_z\,,\quad\cr
 & j_r=\frac1{2\pi}\oint\frac{\sqrt{V_r(r,E,L_z,Q,\mu^2)}}{\Delta(r)} dr\,,\cr
 & j_\theta=\frac1{2\pi}\oint\sqrt{V_\theta(\theta,E,L_z,Q,\mu^2)} d\theta\,. 
\end{align}
Here, $\oint$ means the integral over one cycle of oscillation of the integration variable, and the signature of the square root is chosen to be positive (negative) when the integration variable increases (decreases). 

The motion is dictated for an arbitrary time-parameter by the Hamiltonian~\cite{Kakehi:2024bnh}, 
\begin{align}\label{eq:hamiltoniany}
    H=\frac{y}{2} (g^{\mu\nu} p_\mu p_\nu-1) \,.
\end{align}
Here, $y$ is an undetermined multiplier that is adjusted according to the choice of the time parameter. For proper time, set $y = 1$, and for Mino time, set $y = \Sigma$.
In the case of $g_{\mu\nu}=g_{\mu\nu}^{(0)}$, the equations of motion for $q^\alpha$ and  $j_\alpha$ take the form of
\begin{align}
    \frac{dj_{\alpha}}{d\tau} &= 0\,,\\
    \frac{dq^\alpha}{d\tau} &= \omega^\alpha(j_\beta)\,,
\end{align}
with $\tau$ and $\omega_\alpha$ being the arbitrary time parameter corresponding to the multiplier $y$ and corresponding frequencies, respectively,
which follow immediately from the definition of the action-angle variables.
Namely, the action variables are conserved, and the angular variables evolve uniformly over time.

\subsection{Influence of EMRIs in near region: Resonance}
Here we discuss how DM particles are gravitationally affected by the EMRI secondary. 
Since the mass of the EMRI secondary is much smaller than the central BH, its effect can be treated as a perturbation.  
We denote the perturbed spacetime metric as $g_{\mu\nu}$ and the perturbation due to the EMRI secondary as $h_{\mu\nu}$, {\it i.e.}, 
\begin{align}
g_{\mu\nu}=g_{\mu\nu}^{(0)}+h_{\mu\nu}\,.
\end{align}
Accordingly, the Hamiltonian can be decomposed into background part $H_0$ and the perturbation $H_{\mathrm{int}}$ as
\begin{align}
    H=&H_0+H_{\mathrm{int}}\,,&\\
    H_0:=&\frac{y}{2}\left(g^{\mu\nu}_{(0)}p_\mu p_\nu-1\right)\,,&\\
    H_{\mathrm{int}}:=&-\frac{y}{2}h^{\mu\nu}p_\mu p_\nu\,.&
    \label{eq:interactionH_def}
\end{align}

When the perturbation of the EMRI secondary is added, the motion is not necessarily integrable, and the action variables defined based on the background metric evolve as, 
\begin{align}
 \frac{dj_\alpha}{d\tau}=-\frac{\partial H_{\mathrm{int}}}{\partial q^\alpha}\,.
 \label{eq:Jdot}
\end{align}
Two objects around the BH, the EMRI secondary and the DM particle, interact efficiently only when the tidal resonance condition is satisfied~\cite{Bonga:2019ycj}. 
Since each geodesic orbit has three different frequencies of oscillations corresponding to $(r, \theta, \phi)$ directions, we have six frequencies in total. 
We denote the EMRI and DM frequencies as ${\Omega}^a, {\omega}^a$ ($a=(r, \theta, \phi)$), respectively\footnote{The reason why the $t$-frequency is irrelevant is explained near Eq.~\eqref{eq:resonance_condition} in the next section.}. 

Since the timescale of the orbital motion is much shorter than that of secular evolution of the action variables, we can average the interaction Hamiltonian over the orbital timescale. Furthermore, the oscillatory part of the evolution of the action variables are totally gauge dependent. As discussed in detail in Sec.~\ref{sec:H_int}, the averaged Hamiltonian is given by 
\begin{align}
   \langle  H_{\mathrm{int}}\rangle(\chi) 
   =&\sum_{(n_a,N_a)\in{\mathrm{Res}}}H_{n_a,N_a} \me ^{i \chi} \,, &
   \label{eq:orbital_avr}
\end{align}
where $\chi$ is expressed in terms of the angle variables \(Q^a\) and $q^a$ of the EMRI and DM particles, respectively, as follows:  
\begin{align}
     \chi:= n_a q^a+ N_a Q^a \,,
    \label{eq:resonance_angle}
\end{align}
which does not depend on time for $(n_a,N_a)\in{\mathrm{Res}}$.  
Here, ``Res'' means the set of all combinations of integers $(n_a, N_a)$ that 
satisfy the resonance condition
\begin{align}
    n_a \hat{\omega}^a+N_a \hat{\Omega}^a=0\,,
    \label{eq:rescon}
\end{align}
with
\begin{align}
    \hat{\omega}^a:=\frac{\omega^a}{\omega^t}\,,\qquad \hat{\Omega}^a:=\frac{\Omega^a}{\Omega^t}\,. 
\end{align}

Since the averaged Hamiltonian at resonance depends on the
the angular variables $q^a$ only through the resonant angle $\chi$, 
the following relationship holds (see, e.g.,~\cite{Isoyama:2018sib})
\begin{align}
    \left\langle\frac{d j_a}{d\tau}\right\rangle=-\left\langle\frac{\pa H_{\mathrm{int}}}{\pa q^a}\right\rangle=-n_a \frac{\pa \langle H_{\mathrm{int}} \rangle}{\pa \chi}\,.
\label{eq:averagedjdot}
\end{align}
Using the expression \eqref{eq:orbital_avr}, we can write down the evolution of each action variable as
\begin{align}
     \left\langle\frac{d j_a}{d\tau}\right\rangle=-i\!\!\!\!\sum_{(n_b,N_b)\in{\mathrm{Res}}}\!\!\!\! H_{n_{b},N_b} n_a \me ^{i \chi} \,.
     \label{eq:proportional}
\end{align}
From this expression, it can be seen that the change rate of each action variable is proportional to the corresponding integer $n_a$~\cite{Gupta:2022jdt}.

\begin{figure}[htbp]
    \centering
    \includegraphics[width=85mm]{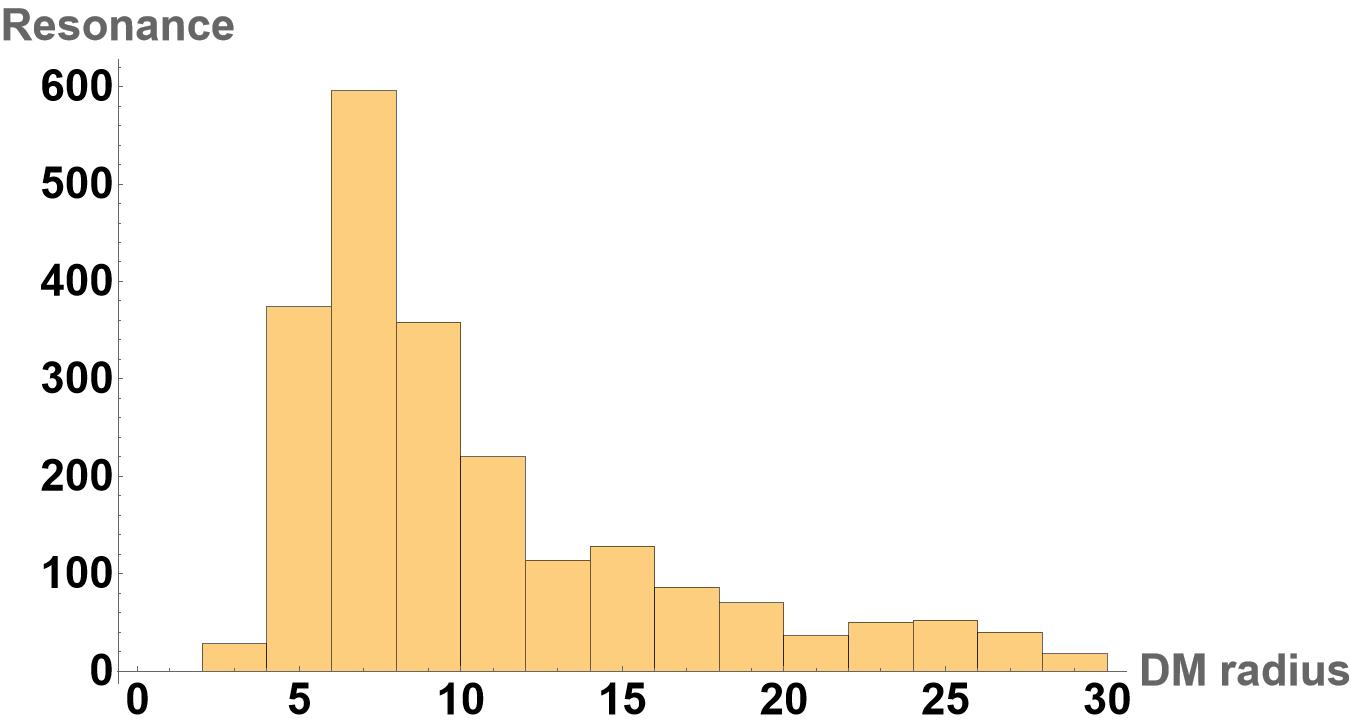}
    \caption{The number of significant resonances experienced by a DM particle in each range of semi-latus rectum $p$ for a fixed EMRI orbit. The horizontal axis represents the semi-latus rectum of the DM particle, normalized by the mass of the central supermassive black hole.
      Each resonance is characterized by six parameters: the eccentricities $e$, the semi-latus recta $p$, and the inclination angles $\iota$ for both DM and EMRI (three parameters each). Instead of $\iota$, we utilize $x=\cos\iota$.
      In this plot, we fix the DM eccentricity and the inclination angles to $(\ec, x) = (0.5, 0.64)$, 
      and the orbital parameters of the EMRI to $(\ec, p, x) = (0.1, 100, 0.7)$.
As the EMRI orbit shrinks due to the GW emission, the inner DM particles hit resonant orbits that shrink accordingly. The vertical axis shows the number of resonances with the sum of the absolute values of resonance integers less than or equal to 10. The size of the bin in the horizontal axis is $2M$. The cumulative number of resonances in this figure is 2231.}
    \label{fig:res1}
\end{figure}

So far, we focused on a single resonance. 
However, once we consider the backreaction of the GW emission, 
the DM particle will experience multiple resonances as the EMRI orbit contracts. 
Here, we discuss the number of resonances experienced by DM particles.
We find that the DM particles in the LISA-band experience a large number of resonances before the EMRI reaches the LISA-band. (see Fig.~\ref{fig:res1}.)

Counting the number of resonances is challenging due to the six parameters involved: the eccentricity, the semi-latus rectum, and the inclination angle $(\ec, p, x)$ of the EMRI system and those of the DM particle. 
In Fig.\ref{fig:res1}, we present the distribution of resonances by setting the DM orbital parameters as follows: eccentricity 
$e=0.5$, inclination $x=0.64$, and varying the semi-latus rectum 
$p$. The EMRI orbital parameters are fixed at $(\ec, p, x) = (0.1, 100, 0.7)$. 
For each bin of the semi-latus rectum of the DM particle, we show the number of resonances with resonance order 
\begin{align}
    K:=\sum_{a}|n_a|+|N_a|\leq 10\,.
\end{align} 
As can be read from the figure, a large number of resonances exist. 
When the EMRI orbit contracts due to the backreaction of GWs, the semi-latus rectum of the DM particle at resonance will also get smaller. As a result, 
all the resonances with the labels $(n_a, N_a)$ located outside of each DM particle in this figure will hit the particle as the EMRI evolves. 
The figure represents just a typical parameter choice, 
but the results for other cases are similar.
Thus, we find that DM particles close to the central BH, in general, experience resonances 
of $O(10^3)$ with $\sum_a(|n_a|+|N_a|)\leq 10$ for each EMRI event.

\subsection{Resonant jump }
Here, we estimate how much the orbital elements change at a single resonance. First, we confirm that the dynamics around a resonance can be treated essentially as a one-dimensional system. 
We assume that the EMRI orbital parameters vary very slowly. The cases in which this 
condition is not satisfied will be discussed in Sec.\ref{sec:fast-evolution}. 

\subsubsection{Reduction to a one-dimensional system}
As noted in the previous section, orbital averaging eliminates contributions 
to $\langle H_{\rm int}\rangle$ from non-resonant Fourier modes. 
As the angle variables other than $\chi$ defined in \eqref{eq:resonance_angle} evolve rapidly, they are irrelevant for long-term evolution. 
To understand the evolution of $\chi$, we only need to know the combination of frequencies ${\cal R}$ defined by
\begin{align}
    \mathcal{R}=n_a \hat{\omega}^a + N_a \hat{\Omega}^a \,,
\end{align}
which is zero exactly at the resonance. 
When we neglect changes in the frequency of the EMRI secondary $\hat{\Omega}^a$ due to the effect of the dissipative self-force, 
we obtain the following equations:
\begin{align}
 \frac{d\chi}{d\tau}=& {\cal R}\,,& \label{eq:dchi_dtau}\\ 
     \frac{d{\cal R}}{d\tau}= &n_a \frac{\partial \hat{\omega}^a}{\partial j_b} \frac{dj_b}{d\tau}\,.&
    \label{eq:eom}
\end{align}
By using the equation of motion \eqref{eq:averagedjdot}, the average of the second equation can be rewritten as
\begin{align}
 \frac{d\langle{\cal R}\rangle}{d\tau}
     =-k \frac{\partial \left\langle H_{\rm int}\right\rangle(\chi)}{\partial \chi}\,,
    \label{eq:dRdtau2}
\end{align}
with 
\begin{align}
k:= G^{ab} n_a n_b\,,\qquad
   G^{ab}:=\frac{\partial \hat\omega^a}{\partial j_b}=\frac{\partial^2 H_0}{\partial j_a \partial j_b}\,.\label{eq:def_Gmetric}
\end{align}

\subsubsection{Motion near the resonance}
The changes in the orbital elements due to resonance can be obtained by solving Eqs.~\eqref{eq:dchi_dtau} and~\eqref{eq:dRdtau2}.
When the interaction Hamiltonian is Fourier decomposed, we expect the fundamental mode to dominate, which leads to the approximation by
\begin{align}\label{eq:Hintapprox}
    \langle H_{\mathrm{int}} \rangle(\chi) =-\mathcal{H} \cos{\chi} \,,
\end{align}
with $\mathcal{H}$ being the constant amplitude of the interaction Hamiltonian. The origin of $\chi$ is chosen such that $\langle H_{\mathrm{int}} \rangle$ takes the minimum there.  

\begin{figure}
    \centering
    \includegraphics[width=\linewidth]{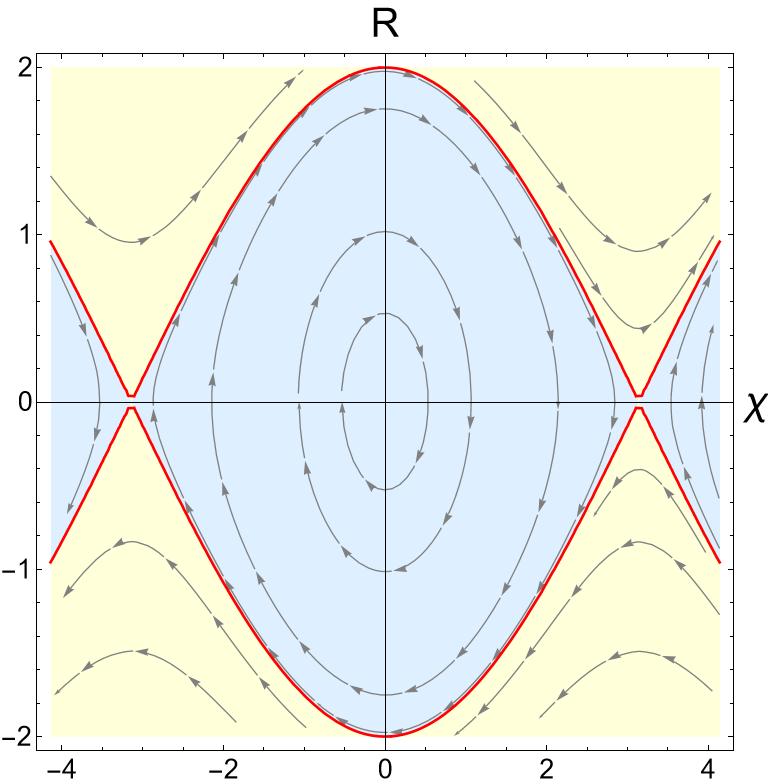}
    \caption{The phase diagram of the motion at a resonance. Similar to a pendulum, the motion is divided into an unbound rotation region (yellow area) and a bound libration region (blue area), with the separatrix indicated by the red curve. When GW backreaction is included, it induces a phase-space flow in the \(\mathcal{R}\)-axis direction, advecting particles so that they traverse the libration region. If the timescale of the GW backreaction is slower than that of the resonance, the shift of the particle can be assumed to occur adiabatically. Since the volume of phase space is conserved, it is expected that particles, except for a few in the libration region, will move in the opposite direction by an amount proportional to the volume of the libration region.}
    \label{fig:resonance_contour}
\end{figure}

Differentiating Eq.~\eqref{eq:dchi_dtau} and substituting Eqs.~\eqref{eq:dRdtau2} and~\eqref{eq:Hintapprox}, the equation of motion for $\chi$ becomes 
\begin{align}
    \frac{d^2 \chi}{d\tau^2}=-\alpha \sin\chi \,,
\end{align}
with 
\begin{align}
 \alpha:=|k \mathcal{H}|\,.
\end{align}
This shows that the resonant angle follows the same dynamics as a pendulum~\cite{murray1999solar}. The motions in the phase space are shown in Fig.~\ref{fig:resonance_contour}.
The maximum amplitude of the oscillation of $\mathcal{R}$ can be easily 
obtained from the “energy conservation” equation, which is given by
\begin{align}
 {\cal R}_{\rm max}=2\sqrt{\alpha}\,.  
 \label{eq:Rmax}
\end{align}

In reality, the GW backreaction is present, and the frequency would gradually drift. The time taken to cross the resonance is roughly given by
\begin{align}
    \Delta \tau_{\text{GW}} := \frac{\mathcal{R}_\max}{d\Omega/d\tau}= &\frac{5}{48} \frac{\sqrt{\alpha}a_{\text{EMRI}}^{\frac{11}{2}} (1 - \ec_{\text{EMRI}}^2)^{\frac{7}{2}}}{\eta M^{\frac{7}{2}}}&\notag \\
    &\times \left( 1 + \frac{73}{24} \ec_{\text{EMRI}}^2 + \frac{37}{96} \ec_{\text{EMRI}}^4 \right)^{-1}~,&\label{eq:delta_GW}
\end{align}
under the lowest order post-Newtonian approximation.
The dynamics would change depending on whether $\Delta \tau_{\text{GW}}$ is slow or not compared to the typical libration timescale, which we call the resonance timescale.
In this work, we set the resonance timescale to the oscillation period when the amplitude of $\chi$ is equal to $\pi$:
\begin{align}
 \Delta \tau_{\text{res}}\sim \frac{4}{\sqrt{\alpha}} K\left(\frac{1}{2}\right)\,, 
 \label{eq:delta_tau}
\end{align}
where $K$ is a complete elliptic integral of the first kind defined by
\begin{align}
    K(k):=&\int_0^{\frac{\pi}{2}} \frac{d\theta}{\sqrt{1-k^2 \sin^2\theta}}\,.
\end{align}

\subsubsection{slow-evolution}

First, let us investigate how the change in the frequency $\mathcal{R}$ is related to the change in the action variables in the case of the slow crossing ($\Delta \tau_{\rm GW} > \Delta \tau_{\rm res}$).
From Eq.~\eqref{eq:proportional}, since the time derivative of each action variable is proportional to the corresponding resonance integer $n_a$, a small jump in the action variable $\Delta j_a$ across a resonance can be written as
\begin{align}
    \Delta j_a =n_a \Theta \,,
    \label{eq:delta_j1}
\end{align}
where $\Theta$ is a proportionality constant and can be determined as follows.
Taking the inner product of equation \eqref{eq:delta_j1} and $n_a$ under metric $G^{ab}$,
\begin{align}
    n^a \Delta j_a=&G^{ab}n_a \Delta j_b&\cr
    =&n_a \Delta \hat\omega^a~.
\end{align}
Here, we use the following relation
\begin{align}
    \Delta\hat\omega^a\simeq G^{ab}\Delta j_b\,,
\end{align}
which can be derived from the definition of $G^{ab}$ \eqref{eq:def_Gmetric}. 
By denoting the average change in $\mathcal{R}$ as $\Delta \mathcal{R}$, we obtain 
\begin{align}
    n_a \Delta\hat\omega^a=&\Delta \mathcal{R}&\cr
    =&k \Theta\,.& \label{eq:Delta_R_j}
\end{align}

Now, our task is to estimate $\Delta \mathcal{R}$. As the EMRI frequencies evolve over time due to GW backreaction, the value of the resonance frequency $\mathcal{R}$ changes.
This change induces a phase space flow opposite to the evolution of \(\mathcal{R}\). Under the influence of this flow, particles are driven along the \(\mathcal{R}\)–axis. Except for a small fraction of particles that remain trapped in the libration region, the vast majority traverse the libration region from one side to the other. By Liouville’s theorem, phase space volume is conserved; hence, the number of particles originally occupying one side of the libration region that cross to the opposite side is proportional to the phase‐space volume of that region~$V$.
This change leads to a shift in the action variables $\Delta j_a$. Since the angle variables are assumed to be randomly distributed on average, 
$\Delta \mathcal{R}$ and the volume of the libration region 
$V$ satisfies the relation
\begin{align}
    \Delta \mathcal{R}=\frac{V}{2\pi}\,.
\end{align}
Since $V$ is the area inside the separatrix, it is calculated as 
\begin{align}
    V=&2\int_0^{2\pi} d\chi \sqrt{2\alpha (1-\sin\chi)}&\cr
    =&16\sqrt{\alpha}\,.&
\end{align}
Thus, $\Delta \mathcal{R}$ is given by
\begin{align}
    \Delta \mathcal{R}=\frac{8\sqrt{\alpha}}{\pi}\,.
\end{align}

Finally, from \eqref{eq:Delta_R_j}, the change in the action variable can be obtained as 
\begin{align}
    \Delta j_a=\frac{8\sqrt{\mathcal{H}}}{\pi |n| }n_a\,.\label{eq:Delta_j}
\end{align}
So far, we have discussed the three action variables \( j_r \), \( j_\theta \), and \( j_\phi \). The remaining \( t \)-component \( j_t\) can be determined by solving the following equation derived from the normalization condition of the four-momentum~\cite{Isoyama:2018sib}, given by
\begin{align}
    \omega^\alpha j_\alpha =-1\,.
\end{align}

From \eqref{eq:Delta_j}, it can be seen that the change in each orbital element is proportional to the square root of the Hamiltonian. 
Since the interaction corresponds to tidal forces in the Newtonian limit, which inversely scale with the cube of the distance from the central BH~\cite{murray1999solar}, the magnitude of the interaction Hamiltonian can be estimated as~\cite{Gupta:2021cno}
\begin{align}
    H_{\text{int}}= \mathfrak{h} \frac{mM^2}{|a_{\text{EMRI}}-a_{\text{DM}}|^3}\,,
    \label{eq:Hint_naive}
\end{align}
where $\mathfrak{h}$ is a dimensionless quantity and is expected to be of $O(1)$ and $a_{\text{DM}}$ is the semi-major axis of a DM particle.
Taking the above into consideration, let us estimate the effects of resonance for a typical EMRI case. We set the mass of the central black hole as $10^6M_\odot$, and consider the case where the mass ratio $\eta=10^{-4}$.
As discussed in the previous section, typically, a DM particle can experience more than $O(10^3)$ resonances. For a mass ratio $\eta\sim O(10^{-4})$ and a radius outside the LISA band $R\sim O(50M)$~\cite{Amaro-Seoane:2012lgq}, each resonance may have an effect of $O(10^{-4})$, since the resonance amplitude scales as the square root of the interaction Hamiltonian.
If all the resonances had the same impact on the orbit, a change in the orbital elements of about $O(1)$ may occur, and the DM orbits may change significantly.
This estimate is too naive because the amplitude of the Hamiltonian involving the high-order resonances should be suppressed. 
In the next section, we will numerically evaluate $\mathcal{H}$ to find that our naive estimate is not always valid.

\subsubsection{fast-evolution}
\label{sec:fast-evolution}
In the above discussion, we assumed that the timescale of oscillation around the resonance point $\Delta \tau_{\text{res}}$ is sufficiently short compared to the timescale of the resonance crossing due to GW backreaction $\Delta \tau_{\text{GW}}$. This assumption might not be always satisfied. 
In such a case, the change in action variables can be approximated by the following equation
\begin{align}
    \Delta j_a\simeq \frac{dj_a}{dt} \Delta \tau_{\text{GW}}\,, \label{eq:delta_j_GW}
\end{align}
and the direction and the amplitude of change in $j_a$ depend on the phase at which the orbit hits the resonance. 

Let us explore the region where $\tau_{\rm GW}$ becomes sufficiently short. Taking the ratio of \eqref{eq:delta_GW} and \eqref{eq:delta_tau} yields
\begin{align}
    \frac{\Delta \tau_{\text{GW}}}{\Delta \tau_{\text{res}}}=&\frac{5}{192 K\left(\frac{1}{2}\right)}\left(\frac{a_{\text{EMRI}}}{a_{\text{EMRI}}-a_{\text{DM}}}\right)^3 a_{\text{EMRI}}^{\frac{5}{2}}\mathfrak{h}k&\notag \\
=&273\left(\frac{a_{\text{EMRI}}}{a_{\text{EMRI}}-a_{\text{DM}}}\right)^3 \left(\frac{a_{\text{EMRI}}}{50M}\right)^{\frac{5}{2}}\mathfrak{h}k&\label{eq:tau_ratio}
\end{align}
Here, we assume the Hamiltonian takes the form of Eq.~\eqref{eq:Hint_naive} and the orbit is circular.
From this result, it can be seen that for most parameter regions in the LISA band, 
the slow-evolution condition, $\Delta\tau_{\rm res} < \Delta\tau_{\rm GW}$, is expected to be satisfied. 
We will see later that this conclusion does not change, even though the estimate of $H_{\rm int}$ given in Eq.~\eqref{eq:Hint_naive} is unreliable.

\section{Derivation of the interaction Hamiltonian}
\label{sec:derivation}
In this section, we derive the expression for the interaction Hamiltonian $H_{\mathrm{int}}$.
We then obtain an explicit expression for the evolution of the orbital elements at a single resonance.
The explicit expression for the evolution of orbital elements during resonance in Kerr spacetime was obtained for the first time by~\cite{Silva:2022blb}. Although their method is very elegant, we derive it in a more formal and easily extendable way. The equivalence of these expressions can be easily shown by using identities given in Ref.~\cite{Sago:2005fn}.

Since the interaction Hamiltonian includes the metric perturbation, we first review the metric reconstruction procedure following Ref.~\cite{Sago:2005fn}. We then average the obtained equations over trajectories to derive the desired expression of the Hamiltonian.

\subsection{Metric reconstruction}
We recapitulate the metric reconstruction from the Teukolsky variable, which satisfies the following separable equation,
\begin{align}
{}_s \mathcal{O} {}_s\Psi=&4\pi \Sigma {}_s \hat{T} \,,&
\label{eq:Teukolsky_eq}
\end{align}
where $s$ represents the spin. Here, ${}_s \mathcal{O}$ is a linear differential operator, which is separable for $r$ and $\theta$ as
\begin{align}
    {}_s \mathcal{O}=&{}_s \mathcal{O}_r+{}_s \mathcal{O}_\theta \,,&
\end{align}
where $\st \mathcal{O}_r$ and $\st \mathcal{O}_\theta$ are represented by only $r$ and $\theta$ coordinates, respectively. Their explicit expressions can be found in Ref.~\cite{Sago:2005fn}.

The source term ${}_s \hat{T}$ on the right-hand side is obtained by acting the linear differential operator $\st\tau_{\mu\nu}$ on the stress-energy tensor $T^{\mu\nu}$ as
\begin{align}
{}_s \hat{T}=&{}_s \tau_{\mu\nu}T^{\mu\nu} \,.&
\label{eq:T_def}
\end{align}
For a point particle, $T^{\mu\nu}$ is given by~\cite{Poisson:2011nh}
\begin{align}
T^{\mu\nu}=m \int_{\gamma} \frac{dz^\mu}{d\tau} \frac{dz^\nu}{d\tau} \delta^{(4)}(x-z(\tau))\,y^{-1}d\tau \,.
\end{align}
Here $\gamma$ is the trajectory of the particle.

The homogeneous solution of the Teukolsky equation is expressed in a mode-decomposed form,
\begin{align}
{}_s \Psi &=\int_{-\infty}^\infty d\omega \sum_{l,m}\me ^{-i\omega t}{}_s R_\Lambda (r) \st Z_\Lambda (\theta,\phi)\,,&
\end{align}
where $\st Z_\Lambda(\theta,\phi)$ is the spin weighted spheroidal harmonics and $\st R_\Lambda(r)$ is the homogeneous solution of the radial equation with $\Lambda=(\omega,l,m)$.
For notational simplicity, we introduce the mode function $\st\Omega_\Lambda^\flat$ defined by
\begin{align}
    \st \Omega_\Lambda^\flat:=\me ^{-i\omega t}{}_s R_\Lambda^\flat (r) \st Z_\Lambda(\theta,\phi)\,,
\end{align}
with $\flat$ being the label specifying the boundary conditions.

Suppose that when $\st \Psi$ is expanded as
\begin{align}
\st \Psi=\sum_\Lambda A_\Lambda \st \Omega_\Lambda\,,
\label{eq:psi_0}
\end{align}
the reconstructed metric $h_{\mu\nu}$ is given by (see Eq.~(A.32) in~\cite{Sago:2005fn})
\begin{align}
h_{\mu\nu}=2 \sum_\Lambda\Re A_\Lambda \st \Pi_{\Lambda,\mu\nu}\,,
\label{eq:reconstruction_0}
\end{align}
where
\begin{align}
\st \Pi^\flat_{\Lambda,\mu\nu}:=&\zeta_s \st\tau^\ast_{\mu\nu}\st\tilde{\Omega}^\flat_\Lambda\,,&\\
\st \tilde{\Omega}_\Lambda^\flat:=&{}_{-s}R_\Lambda^\flat \st Z_\Lambda \me ^{-i\omega t}\,,&\label{eq:Omega_tilde_def}
\end{align}
and $\Re$ denotes the real part.
Here, we define the adjoint operator $\st\tau_{\mu\nu}^\ast$ so as to satisfy
\begin{align}
    \int d^4x \sqrt{-g}\, \overline{X} \st \tau_{\mu\nu} Y^{\mu\nu}= \int d^4x  \sqrt{-g}\, Y^{\mu\nu}\, \overline{\st \tau_{\mu\nu}^\ast X}\,,
    \label{eq:adjoint_def}
\end{align}
for an arbitrary scalar field $X$ and a tensor field $Y^{\mu\nu}$ which vanish on the boundaries. 
Here, $\overline{X}$ is the complex conjugate of $X$.
The constant $\zeta \st$ is related to the Teukolsky-Starobinsky constant $\mathcal{C}$~\cite{1974ApJ...193..443T} as 
\begin{align}
    \zeta_2=\frac{1}{\mathcal{C}}\,,\qquad
    \zeta_{-2}=\frac{16}{\overline{\mathcal{C}}}\,.
\end{align}
From above, the metric perturbation can be obtained by calculating the coefficients $A_\Lambda$ by solving the Teukolsky equation \eqref{eq:Teukolsky_eq}. 
In the following, we discuss how to obtain the coefficients $A_\Lambda$, specifically for the case of a point particle.

The Green's function of the Teukolsky equation \eqref{eq:Teukolsky_eq} $\st G(x,x^\prime)$, in general, satisfies
\begin{align}
\st \mathcal{O} \st G(x,x^\prime)=\frac{\delta^{(4)}(x-x^\prime)}{\Delta^s} \,. 
\label{eq:Green_def}
\end{align}
When the source radius $r'$ is greater than that of the field point $r$, the Green’s function with the retarded boundary condition is simply given by
\begin{align}
\st G(x,x^\prime)=\sum_\Lambda \frac{1}{2\pi W_\Lambda}\st \Omega^{\mathrm{in}}_\Lambda(x) \st R^{\mathrm{up}}_\Lambda (r^\prime)\overline{\st Z_\Lambda (\theta^\prime,\phi^\prime) }\me ^{i\omega t^\prime}\,, 
\label{eq:Green_0}
\end{align}
where $W$ is the Wronskian between$\st R^{\mathrm{in}}_\Lambda$ and $\st R^{\mathrm{up}}_\Lambda$ defined by
\begin{align}
    W_\Lambda:=\Delta^{s+1}\left[\st R^{\mathrm{in}}_\Lambda \frac{d}{dr}\st R^{\mathrm{up}}_\Lambda-\st R^{\mathrm{up}}_\Lambda \frac{d}{dr}\st R^{\mathrm{in}}_\Lambda\right]\,.
\end{align}
Here, $\st R^{\mathrm{up}}_\Lambda$ and $\st R^{\mathrm{in}}_\Lambda$ are the homogeneous solutions which satisfy the following respective boundary conditions,
\begin{align}
\st R_{\Lambda}^{\text{in}} &:= 
\begin{cases} 
\st B_{\Lambda}^{\text{inc}} r^{-1} e^{-i \omega r^*} + \st B_{\Lambda}^{\text{ref}} r^{-2s-1} e^{i \omega r^*}, &  r^* \to \infty, \\
\st B_{\Lambda}^{\text{trans}} \Delta^{-s} e^{-ikr}, & \!\!\!\! r^* \to -\infty,
\end{cases} \\
\st R_{\Lambda}^{\text{up}} &:= 
\begin{cases} 
\st C_{\Lambda}^{\text{trans}} r^{-2s-1} e^{-i \omega r^*}, &  r^* \to \infty, \\
\st C_{\Lambda}^{\text{up}} e^{ikr} + \st C_{\Lambda}^{\text{ref}} \Delta^{-s} e^{-ikr}, & r^* \to -\infty,
\end{cases} \
\end{align}
with $r^\ast:=\int dr (r^2+a^2)/\Delta$ being the tortoise coordinate.

We also define $\st R^{\mathrm{down}}_\Lambda$ to satisfy
\begin{align}
{}_{s}R^{\mathrm{up}}_\Lambda:=\Delta^{-s} \overline{{}_{-s} R^{\mathrm{down}}_\Lambda}\,.
\label{eq:down_def}
\end{align}
It is known that $\st R^{\text{up}}_\Lambda$ thus obtained from $\st R^{\text{down}}_\Lambda$ gives a homogeneous solution of the Eq.~\ref{eq:Teukolsky_eq}.
Substituting \eqref{eq:down_def} into \eqref{eq:Green_0},
we obtain the following equation
\begin{align}
\st G(x,x^\prime)=\sum_\Lambda \frac{1}{2\pi W_\Lambda}\st \Omega^{\mathrm{in}}_\Lambda(x) \overline{{\st\tilde{\Omega}^{\mathrm{down}}_\Lambda}}(x^\prime) \Delta^{-s}(r^\prime)\,,
\label{eq:Green_1}
\end{align}
using the definition \eqref{eq:Omega_tilde_def}.

From Eqs.~\eqref{eq:Green_def} and \eqref{eq:Green_1}, we can write down the solution to Eq.~\eqref{eq:Teukolsky_eq} as
\begin{align}
\st \Psi=&\int d^4x^\prime G(x,x^\prime)\Delta^s 4\pi \Sigma \st\hat{T}\sin\theta^\prime&\cr
=&\sum_\Lambda \st \Omega^{\mathrm{in}}_\Lambda(x) \frac{ 4 }{W_\Lambda \bar{\zeta_s}}\int d\tau^\prime y\overline{\st \Pi_{\Lambda,\mu\nu}^{\mathrm{down}}}p^\mu p^\nu\,, &
\label{eq:psi_1}
\end{align}
where we perform integration by parts and use Eqs.~\eqref{eq:T_def} and \eqref{eq:adjoint_def}. We also use the fact that the determinant of the metric is given by
\begin{align}
    \sqrt{-g}=\Sigma \sin\theta\,.
\end{align}

By comparing Eqs.~\eqref{eq:psi_0} and \eqref{eq:psi_1}, the coefficient $A_\Lambda$ is determined as,
\begin{align}
A_\Lambda= \frac{ 4 }{W_\Lambda \bar{\zeta_s}}\int d\tau^\prime y \overline{\st \Pi_{\Lambda,\mu\nu}^{\mathrm{down}}}p^\mu p^\nu\,.
\end{align}
Substituting this into Eq.~\eqref{eq:reconstruction_0}, the desired expression of the reconstructed metric $h_{\mu\nu}(x)$ is given by
\begin{align}
h_{\mu\nu}=\Re \sum_\Lambda \frac{8 }{W_\Lambda \bar{\zeta_s}} \st \Pi^{\mathrm{in}}_{\Lambda,\mu\nu} \int d\tau^\prime y\overline{\st \Pi_{\Lambda,\alpha\beta}^{\mathrm{down}}}p^\alpha p^\beta\,,
\label{eq:metric_fin}
\end{align}
unless the source is extended down to the maximum radius of the DM orbit.

\subsection{Interaction Hamiltonian}
\label{sec:H_int}
Now, we derive the analytic expression of the interaction Hamiltonian.
Since the timescale of the change in the action variables is much longer than the timescale of the orbital oscillations, the interaction Hamiltonian can be averaged over time.
Plugging Eq.~\eqref{eq:metric_fin} into Eq.~\eqref{eq:interactionH_def} and taking the time average, we obtain
\begin{align}
\langle H_{\mathrm{int}}\rangle=-&\lim_{T\to\infty}\frac{1}{2T}\int_{-T}^T d\tau \frac{y}{2}h^{\mu\nu}u_\mu u_\nu&\cr
=\lim_{T\to\infty}&\frac{1}{2T}\Re \sum_\Lambda \frac{ 4}{W_\Lambda \bar{\zeta_s}}\int_{-T}^T d\tau 
\psi^{\mathrm{in}}_\Lambda 
 \int_\gamma d\tau^\prime \overline{\psi^{\mathrm{down}}_\Lambda}\,, &
 \label{eq:Hint_avr1}
\end{align}
where 
\begin{align}
\psi_\Lambda^\flat:=&y \Pi^\flat_{\Lambda,\mu\nu} p^\mu p^\nu\,.
\end{align}
Note that $\psi^{\mr{in}}$ is evaluated along the inner DM particle orbit, while $\psi^{\mr{down}}$ along the outer EMRI secondary orbit.

Due to the periodicities of $r$ and $\theta$, we expand $\psi_\Lambda^\flat$ into a Fourier series with respect to the angular variables $q^\alpha$:
\begin{align}
\psi_\Lambda^\flat=&\sum_{k,n}\psi^\flat_{\Lambda,kn}\me ^{i\left(\omega \omega^t-m\omega^\phi-n\omega^r-k\omega^\theta\right)\tau+i q^0}\,,
\label{eq:psi_fourier}
\end{align}
where the Fourier coefficients are given by
\begin{align}
\psi_{\Lambda,kn}^\flat=& \frac{1}{4\pi^2}\int dq^r dq^\theta \psi^\flat_\Lambda(r(q^r,q^\theta),\theta(q^r,q^\theta))\me ^{i( n q^r+k q^\theta)}~.
\end{align}
Taking the time integral of Eq.~\eqref{eq:psi_fourier}, we obtain
\begin{align}
\int d\tau\psi_\Lambda^{\mathrm{down}}=\sum_{k,n}\psi_{\Lambda,kn}^{\mathrm{down}}\frac{2\pi}{\Omega^t}\delta\left(\omega-N_a \frac{\Omega^a}{\Omega^t}\right)\me ^{i Q^0}\,.
\label{eq:psi_avr}
\end{align}
Substituting Eqs.~\eqref{eq:psi_fourier} and \eqref{eq:psi_avr} into Eq.~\eqref{eq:Hint_avr1} and performing the $\omega$-integral, we obtain 
\begin{align}
    \langle H_{\mathrm{int}}\rangle=&\sum_{lm}\frac{4}{\bar{\zeta_s} W_{lm}}\sum_{k^\prime,n^\prime}\sum_{k,n}\frac{2\pi}{\Omega^t}\psi^{\mr{in}}_{lm,k^\prime n^\prime}\overline{\psi^{\mr{down}}_{lm,kn}}\cr
&\times\me ^{iq^0-iQ^0}\lim_{T\to\infty}\frac{1}{2T}\int_{-T}^T d\tau\me ^{i\left(\frac{\omega^t}{\Omega^t}N_a \Omega^a-n_a \omega^a\right)\tau}\,.\cr
\end{align}
Taking the time average, we obtain the analytic expression for the averaged interaction Hamiltonian as 
\begin{align}
    \langle H_{\mathrm{int}}\rangle=&\sum_l\frac{8\pi}{\bar{\zeta_s} }\sum_{\rm{Res}}\frac{\me ^{iq^0-iQ^0}}{W_{lm}\Omega^t}\psi^{\mr{in}}_{lm,k^\prime n^\prime}\overline{\psi^{\mr{down}}_{lm,kn}}\,,
\end{align}
where $\sum_{m,k,n,k^\prime,n^\prime \in \rm{Res}}$ sums over $(m,k^\prime,n^\prime,k,n)$ satisfying the resonance condition given by
\begin{align}
&\frac{N_a \Omega^a}{\Omega^t}=\frac{n_a \omega^a}{\omega^t}\,.
\label{eq:resonance_condition}
\end{align}
It is more convenient to use the Mino-time instead of the proper time for the numerical computation. The relation between the Hamiltonian in the Mino-time and the proper time can be obtained with the method in~\cite{Kakehi:2024bnh}.

\section{application: resonant scattering of a DM particle}
\label{sec:result}
In this section, we discuss the numerical results of the orbital evolution of DM particles due to multiple resonances.

We treat the time evolution of DM particles as a restricted three-body problem with an EMRI. The orbit of the DM particle is treated as a test particle. The orbital parameters are modified each time when it satisfies the resonance condition, according to Eq.~\eqref{eq:Delta_j} or Eq.~\eqref{eq:delta_j_GW}, depending on whether the ratio~\eqref{eq:tau_ratio} exceeds unity or not. 

To evaluate the latter equation, we adopt a prescription that selects the phase to maximize the contribution from the resonance.
The EMRI orbit is evolved by the GW backreaction using the quadrupole formula. For simplicity, we neglect the influence of DM on the EMRI orbit. 
As mentioned above, we use Mino time by setting
$y = \Sigma$.
All numerical calculations are performed using {\it Mathematica} and Julia, with the Teukolsky mode functions computed by Black Hole Perturbation Toolkit~\cite{BHPToolkit}. In the previous section, we demonstrate the existence of the conserved quantities \((\mu^2,E, L_z, Q)\). However, it is customary to describe bound trajectories using the orbital elements—the semi-latus rectum \(p\), eccentricity \(e\), and inclination \(x\)—and we adopt this convention throughout this section. The relation between \((\mu^2,E, L, Q)\) and \((p, e, x)\) is described in \cite{Fujita:2009bp}.
In the following calculation, we fix a central BH spin to $a = 0.9$ and the mass ratio to $\eta = 10^{-4}$.

\begin{figure}[htbp]
    \centering    \includegraphics[width=\linewidth]{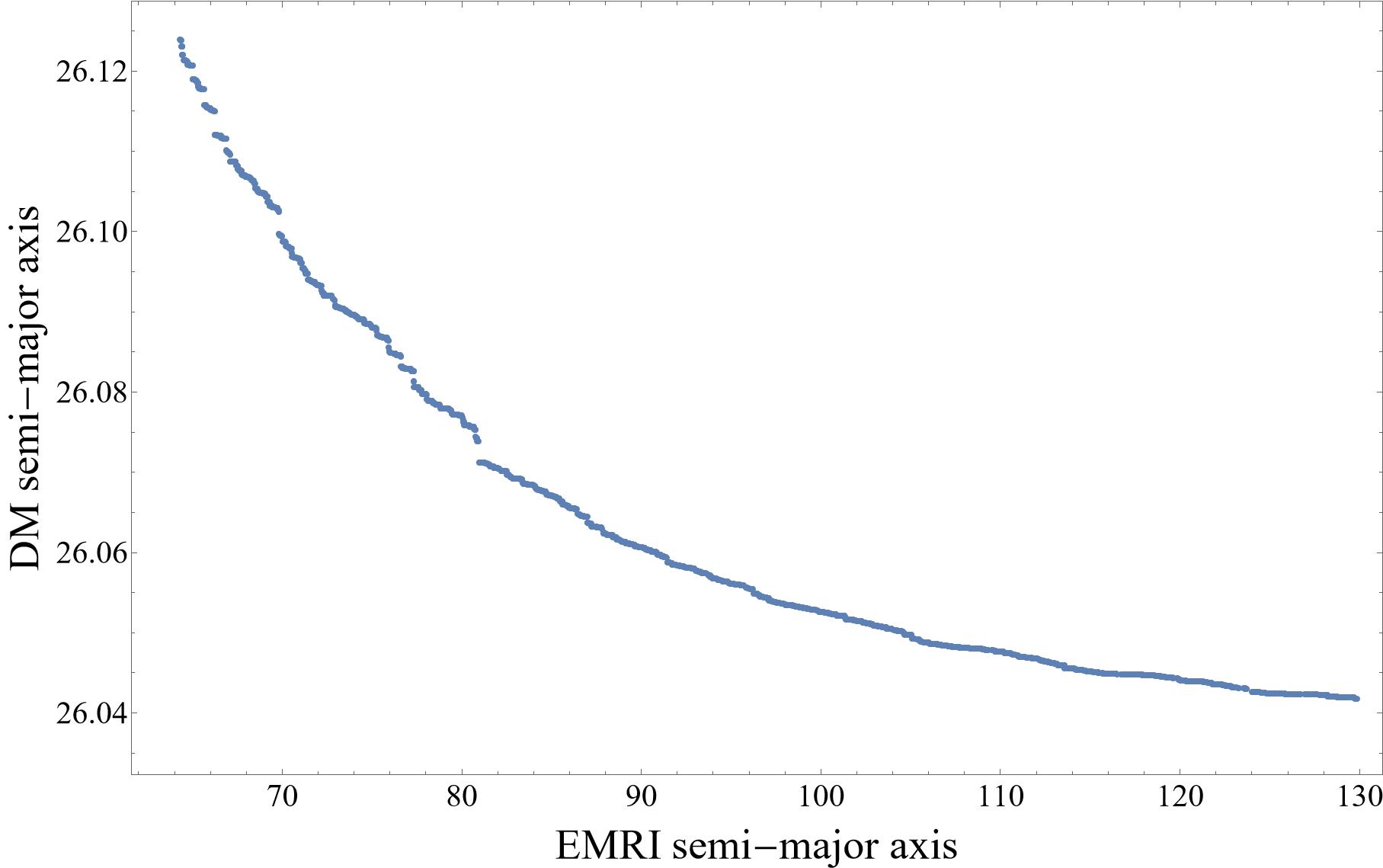}
    \caption{
    The semi-major axis of the DM particle and the EMRI secondary just before the respective resonances. The horizontal and vertical axes correspond to the semi-major axis of EMRI and DM in the unit of the central BH mass $M$, respectively. The jumps in the vertical direction represent the orbital change of the DM particle across the resonance. Since the orbit of the EMRI secondary shrinks due to GW emission, time lapses from right to left. In this figure, \resnum resonances are taken into account, and the total change in semi-major axis is about $\deltaa\%$. The initial orbital parameters of DM particle and EMRI secondary are $(a,p_{\text{DM}},\ec_{\text{DM}},x_{\text{DM}})=(0.9,\pDM,\ecDM,\xDM)$, and $(p_{\text{EMRI}},\ec_{\text{EMRI}},x_{\text{EMRI}})=(\pE,\ecE,\xE)$, respectively. We fixed the mass ratio to $\eta=10^{-4}$.
    For fast resonances, the initial phase was chosen to maximize the change in angular momentum, resulting in a decrease of the DM’s angular momentum.}
    \label{fig:pDM_pEMRI.jpg}
\end{figure}

Figure~\ref{fig:pDM_pEMRI.jpg} shows the evolution of the semi-major axis of the DM particle and that of the EMRI secondary for a typical set of initial orbital parameters\footnote{In the Newtonian limit, the semi–major axis is directly related to the orbital energy~\cite{murray1999solar}, making it a natural measure of the orbit’s size; accordingly, we plot the change in the semi–major axis.}. Each point shows the semi-major axis of the DM particle (vertical direction) and that of EMRI (horizontal direction) just before the resonances. 
As the EMRI orbit shrinks, the state evolves from right to left in this figure. We terminate our numerical calculations when the EMRI secondary orbit overlaps the DM orbit. Note that our current formulation cannot handle situations where these two orbits overlap in $r$. In this figure, the evolution of the EMRI secondary was computed from the semi-major axis $a_{\text{EMRI}}=\aE M$ (semi-latus rectum $p_{\text{EMRI}}=\pE M$) to $a_{\text{EMRI}} = \aEf M$ ($p_{\text{EMRI}}=\pEf M$), where $M$ is the mass of the central BH. There are \resnum resonances in this figure. As is evident from the vertical axis of the figure, the orbital radius of the DM particle changes little. In this case, the total change of semi-major axis is about $\deltaa\%$. 
We find that only a few resonances contribute significantly and that the DM radius did not change in most of the resonances.

\begin{figure}\centering\includegraphics[width=\linewidth]{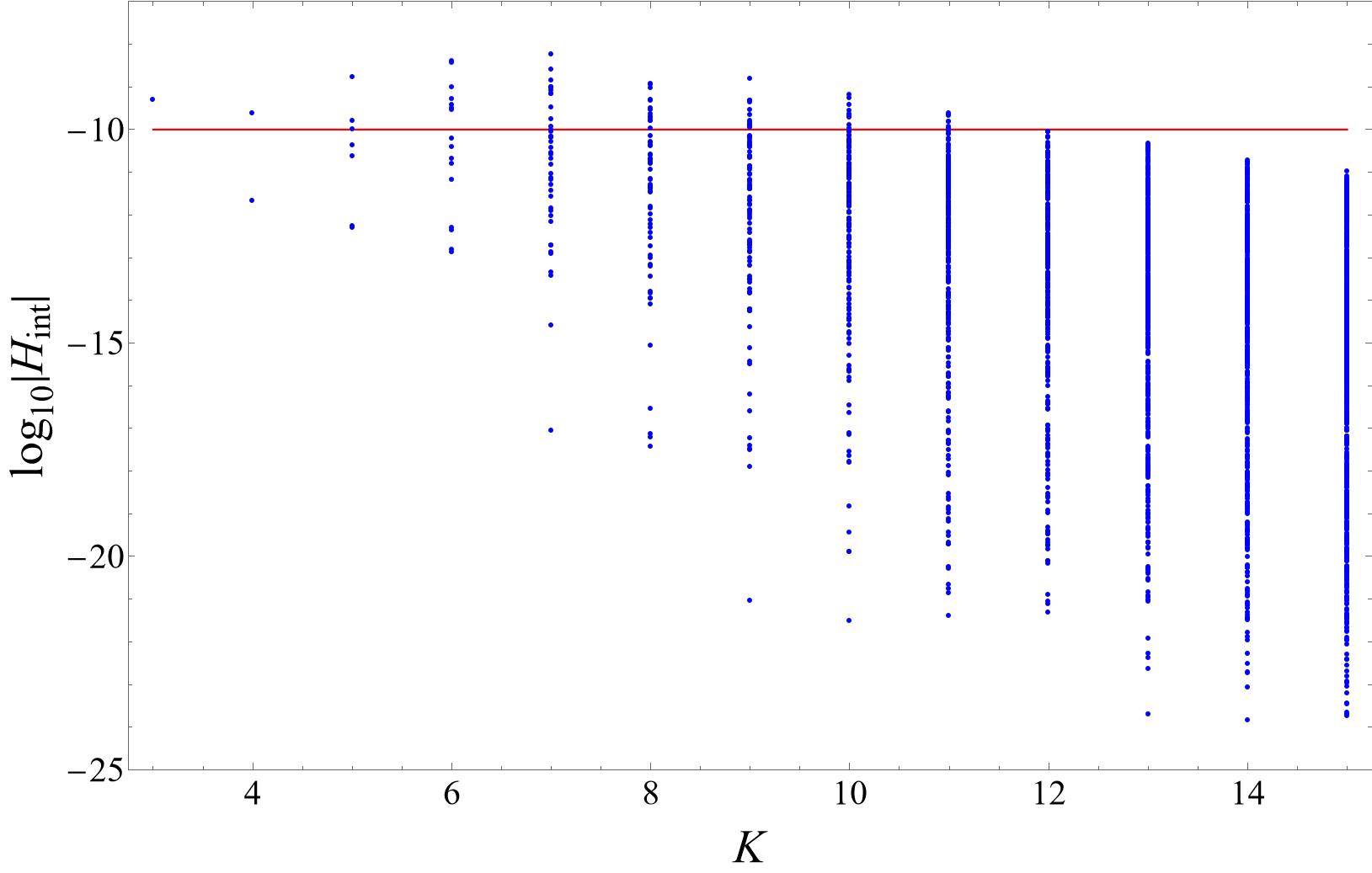}
    \caption{The relationship between the interaction Hamiltonian and the resonance order $K:=\sum_{a}|n_a|+|N_a|$. The blue points represent the absolute values of the interaction Hamiltonian at each resonance. The red line indicates the naively estimated magnitude of the interaction Hamiltonian \eqref{eq:Hint_naive} $\eta/100^3\sim10^{-10}$. The initial parameters are the same as those in Fig.~\ref{fig:pDM_pEMRI.jpg}. It can be read that the interaction Hamiltonian decays exponentially with respect to the resonance integers.}
    \label{fig:|Hint|}
\end{figure}

\begin{table}[bh!]
\centering
\caption{The ratio of semi-major axis between the initial and the final time for various values of the initial eccentricity and orbital inclination angle of the DM particle. The other parameters are fixed in the same way as in Fig.~\ref{fig:pDM_pEMRI.jpg}. 
The semi-major axis changes by at most approximately \deltaamax$\%$, 
indicating that the orbit remains unchanged. 
Additionally, it can be observed that larger eccentricities and inclination angles result in greater orbital change.}
\label{table:redm_xdm}
\begin{tabular}{|c|c|c|c|}
\toprule
$\ec_{\mathrm{DM}} \backslash x_{\mathrm{DM}}$ & $0.7$ & $0.8$ & $0.9$ \\
\hline
0.2 & 1.003 & 1.003 & 1.002 \\
0.3 & 1.005 & 1.004 & 1.004 \\
0.4 & 1.008 & 1.007 & 1.007 \\
\hline\hline
\end{tabular}
\vspace{2mm}
\caption{The ratio of the eccentricities as shown for the semi-latus rectum in Table \ref{table:redm_xdm}. The eccentricity decreases owing to the resonance. The orbit of dark matter (DM) gains angular momentum and becomes circularized, but the effect is small.
}
\begin{tabular}{|c|c|c|c|}
\toprule
$\ec_{\mathrm{DM}} \backslash x_{\mathrm{DM}}$ & $0.7$ & $0.8$ & $0.9$ \\
\hline
0.2 & 0.955  & 0.968  & 0.985  \\
0.3 & 0.971  & 0.981  & 0.990  \\
0.4 & 0.977  & 0.985  & 0.993  \\
\hline\hline
\end{tabular}
\label{table:edm_xdm}
\end{table}

The results show that the effects of resonances are weaker than our previous naive estimate. 
To investigate the reason for the smallness, we plotted the absolute value of the interaction Hamiltonian against the resonance order $K$, which is defined 
 as the sum of the absolute value of the resonance integers, $K:=\sum_{a}|n_a|+|N_a|$, in Fig.~\ref{fig:|Hint|}. 
 Each point shows the magnitude of the Hamiltonian for each resonance, with the horizontal axis representing the resonance order $K$.
Given the mass ratio of $10^{-4}$ and the distance of the EMRI being approximately $100M$, we can estimate that the magnitude of the Hamiltonian in the Newtonian limit is about $10^{-10}$ using Eq.~\eqref{eq:Hint_naive}. 
 From Fig.~\ref{fig:|Hint|}, we observe that the magnitude of the Hamiltonian decays exponentially as the resonance order $K$ increases. 
As higher-order resonances are exponentially suppressed,  
the accumulated effect is smaller than naively expected.
This tendency is expected from the general fact that the higher-order Fourier coefficients of a smooth function decay exponentially. 
 It is not surprising that the Hamiltonian asymptotically decays in the limit 
 of a large resonance order. 
 However, 
 where the expected exponential decay begins can only be determined through actual calculation. 
 In this study, it was confirmed that the decay of the Hamiltonian begins at a small value of $K$. 
 Furthermore, even for the same value of $K$, the magnitude of the interaction Hamiltonian varies by six orders of magnitude. 
 As a result, only a small number of lower-order resonances affect the orbits of DM particles, and the cumulative effect of resonances remains small.

\begin{figure}[htbp]
    \centering
    \includegraphics[width=\linewidth]{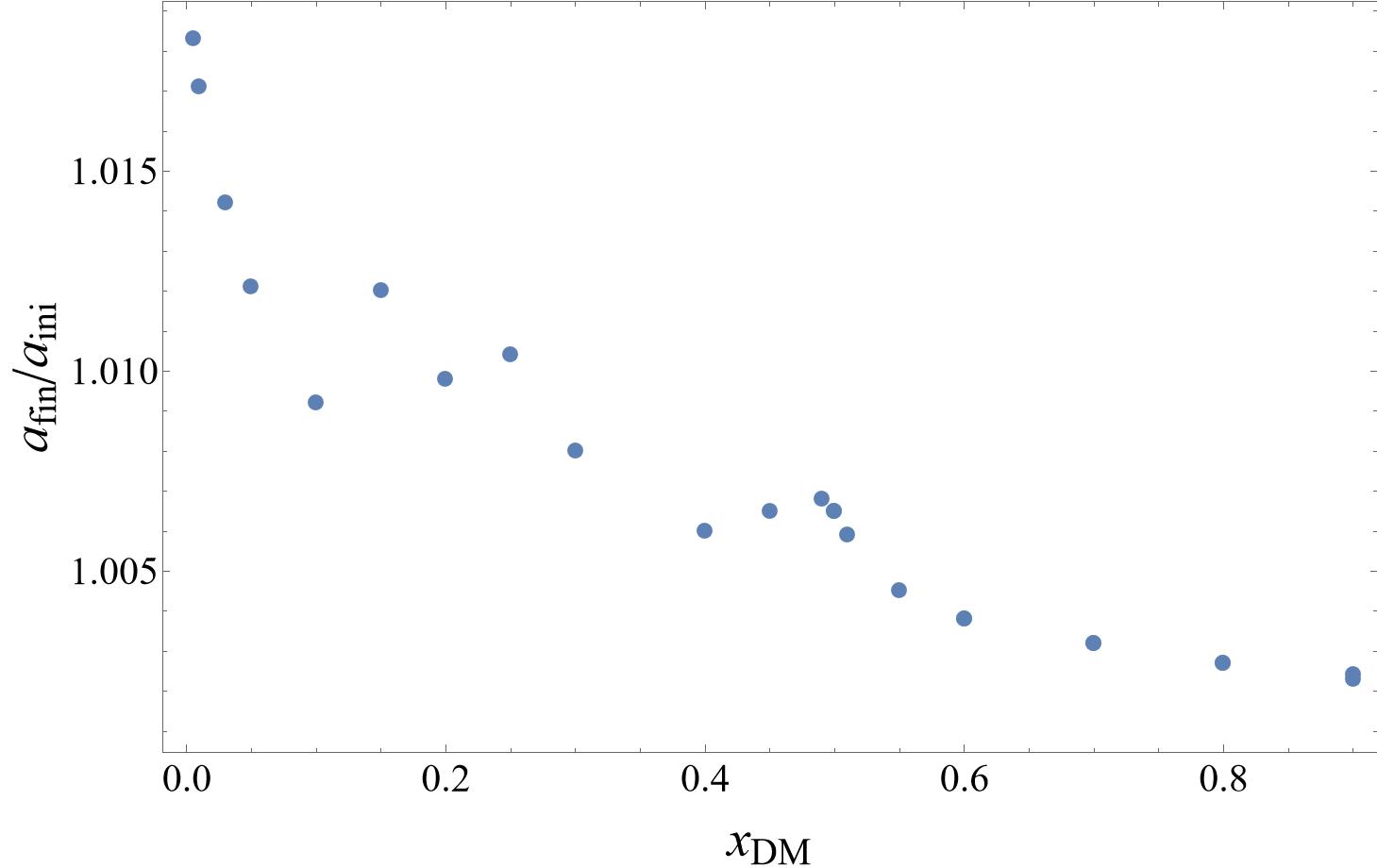}
    \caption{Total change in the dark-matter particle’s semi-major axis as a function of \(x_{\rm DM}\). The horizontal axis corresponds to \(x_{\rm DM}\), while the vertical axis shows the cumulative fractional change in the semi-major axis. Although the variation grows as \(x_{\rm DM}\) decreases, it remains at the percent level ($\sim$1\%). All other parameters are the same as those used in Fig.~\ref{fig:pDM_pEMRI.jpg}.}
\label{fig:xDM_aratio}
\end{figure}

Tables~\ref{table:redm_xdm} and~\ref{table:edm_xdm}, respectively, show the ratios of the semi-major axes and eccentricities between the initial time and the final time for various values of DM eccentricity and the inclination. The other parameters for DM and EMRI are unchanged from Fig.~\ref{fig:pDM_pEMRI.jpg}. 
As can be seen from the table, the cumulative effect of the resonances is small regardless of the parameters. We can observe a trend where larger eccentricities and inclination angles result in greater changes in the orbit. 
It should be noted that the calculation is stopped before the orbital radii of the EMRI and DM start to overlap. When the eccentricity is large, orbital crossing occurs earlier, resulting in a smaller number of computed resonances and potentially underestimating the effect.
We also examined orbits with smaller values of \(x\)—that is, with larger inclination—than those listed in the table. The results are shown in Fig.~\ref{fig:xDM_aratio}. Although the cumulative effect of the resonance grows with inclination, it remains too small to produce a significant change in the orbit of the DM particle. Moreover, for high-inclination trajectories, the interaction Hamiltonian \(H_{\rm int}\) decreases more slowly as \(K\) increases than in the low-inclination case, as detailed in the Appendix~A. This behavior degrades convergence with respect to \(K\) and hence raises the computational cost. 

\begin{figure}[htbp]
    \centering
    \includegraphics[width=\linewidth]{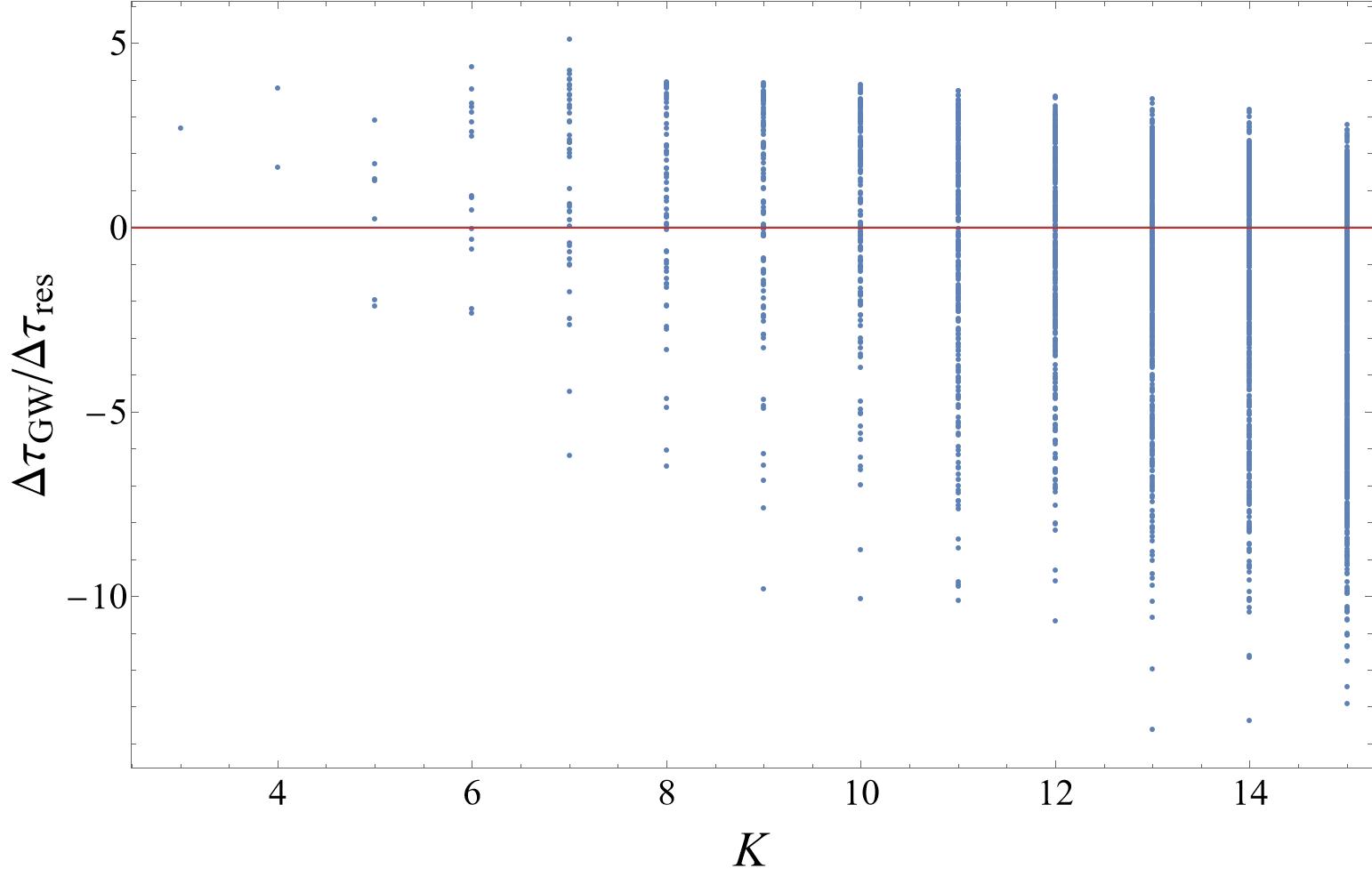}
    \caption{The ratio of the timescales between gravitational back-reaction and resonance $\Delta\tau_{\text{GW}}/\Delta \tau_{\text{res}}$ (see Eq.~\eqref{eq:tau_ratio}). We fixed the initial parameters to those in Fig.~\ref{fig:pDM_pEMRI.jpg}. The horizontal axis represents the resonance integer \( K \), and the vertical axis represents the ratio $\Delta\tau_{\text{GW}}/\Delta \tau_{\text{res}}$. The red line indicates where the ratio equals one. Above the red line, the timescale associated with GWs is sufficiently long, and the orbital changes due to resonance are considered adiabatic. Out of \resnum resonances, \resscale (\tauratio$\%$) lie above the red line. As discussed in Fig.~\ref{fig:|Hint|}, as the resonance integer \( K \) increases, the Hamiltonian decays exponentially, making resonances with larger integers more likely to be non-adiabatic.}
\label{fig:tau_ratio}
\end{figure}

Figure.~\ref{fig:tau_ratio} shows the numerically evaluated ratio between $\Delta\tau_{\rm res}$ and $\Delta\tau_{\rm GW}$ discussed in Eq.~\eqref{eq:tau_ratio}. 
The horizontal axis represents the resonance integer \( K \), 
while the vertical axis represents the ratio $\Delta\tau_{\text{GW}}/{\Delta\tau_{\text{res}}}$. 
A red horizontal line indicates where the ratio equals to one. 
The resonances above the red line belong to the slow-evolution,  
which account for about \tauratio\% of all resonances in Fig.~\ref{fig:tau_ratio}. 
As previously mentioned, when the resonance integer \( K \) increases, the Hamiltonian decays exponentially, leading to the fast-evolution regime. 

The analysis presented thus far has considered fixed EMRI parameters; however, systematic parameter variation studies reveal negligible quantitative differences in the overall results. Complete numerical datasets supporting these findings will be archived in a publicly accessible repository \href{https://sites.google.com/view/bhpc1996/home}{https://sites.google.com/view/bhpc1996/home}.

\section{conclusion and discussion}
\label{sec:conclusion}

In this study, we investigated the effect of resonances on the DM spike during the orbital evolution of EMRI. Approximating the DM particle as a test particle, we formulated the orbital evolution of DM particles as a restricted three-body problem. While the effect of each individual resonance is expected to be small, the cumulative effect of numerous resonances could potentially lead to significant changes in the trajectory. 
To examine this idea, we derived an explicit expression for the interaction Hamiltonian and numerically evaluated the cumulative effect of the resonances. 
We found that the effect on the overall effect on the DM orbital evolution remains very small regardless of the initial orbits of the DM particles, even if the effects of multiple resonances are taken into account. This is because only a small number of the resonances are relevant since the magnitude of the Hamiltonian decays exponentially as the resonance integers increase (see Fig.~\ref{fig:|Hint|}). 
Although the decay of the magnitude of the resonant Hamiltonian can be understood as a result of phase cancellation, the magnitude scatters by about six orders of magnitude even for the same value of the sum of the resonance integers. 
The origin of this scatter of the magnitude of the resonance Hamiltonian is an open question for future work.

In addition, the present study does not take into account the case where the radial region of the DM particle motion and 
that of the EMRI secondary overlap. 
In the case of this radial overlapping, the expression of the Hamiltonian derived in this study cannot be used. 
Our derivation relies on the metric reconstruction scheme assuming a vacuum. To handle the radial overlapping, a metric reconstruction with matter must be used. Intuitively, we expect that the DM particle and the EMRI secondary will get closer to each other in this case, and thus the effect of resonance will be enhanced. Incorporating this effect is also left for future work.

\begin{acknowledgments}
T. Kakehi is supported by JST SPRING, Grant Number JPMJSP2110.
T. Takahashi is supported by JSPS KAKENHI Grant Nos. JP23KJ1214, JP25KJ0067, and JP25K17397.
T.Tanaka is supported by JSPS KAKENHI Grant Nos. JP23H00110, JP24H00963, and JP24H01809. 
H. Omiya is supported by JSPS KAKENHI Grant Numbers JP23H00110 and Yamada Science Foundation. 
\end{acknowledgments}

\appendix

\section{numerical convergence}
\begin{figure}[htbp]
  \centering
\begin{subfigure}{0.48\textwidth}
  \centering
  \includegraphics[width=\linewidth]{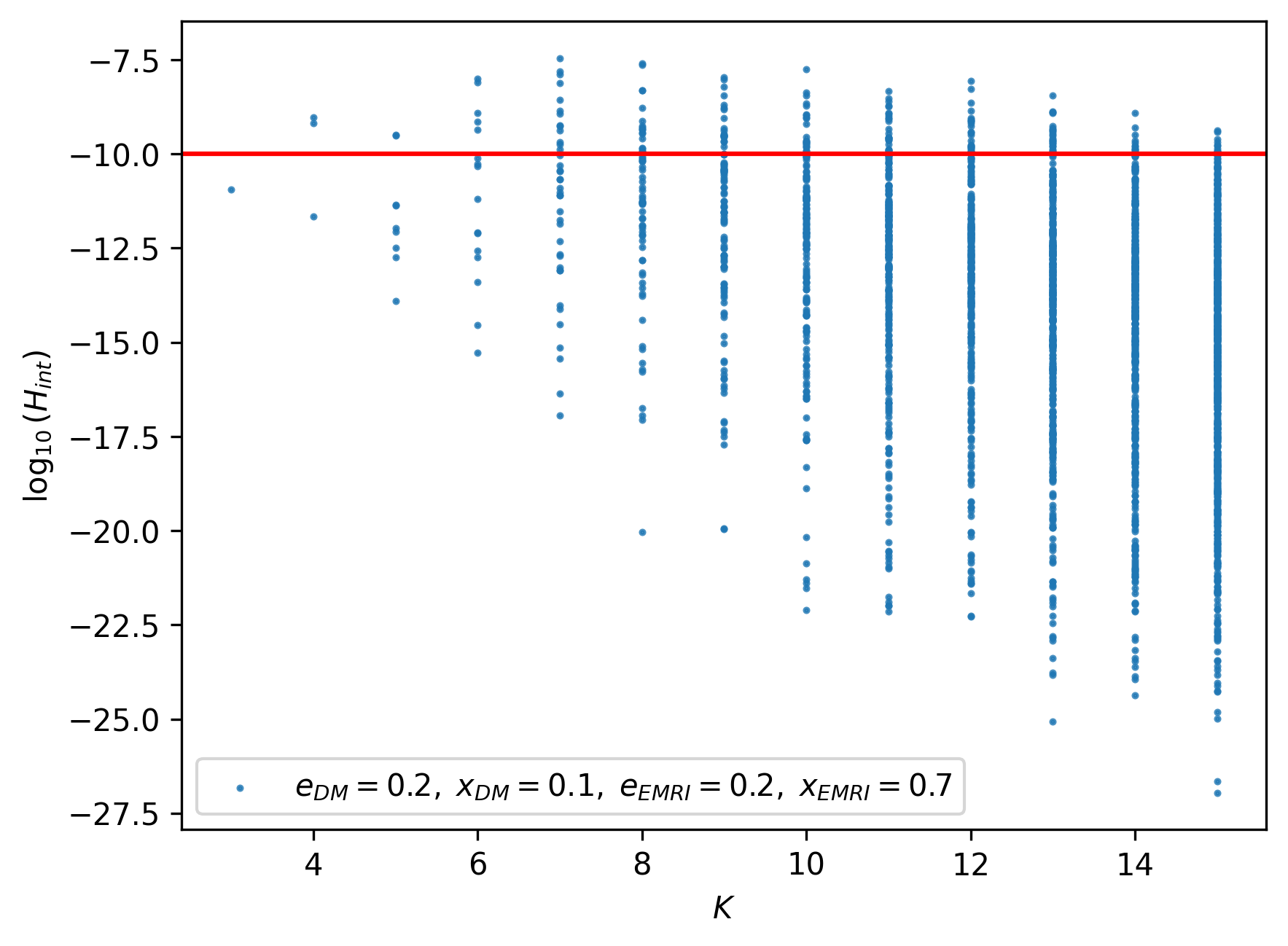}
  \caption{\(x_{\rm DM}=0.1\)}
  \label{fig:hint_m_eE0.2_xE0.7_eDM0.2_xDM0.1.png}
\end{subfigure}
\begin{subfigure}{0.48\textwidth}
  \centering
  \includegraphics[width=\linewidth]{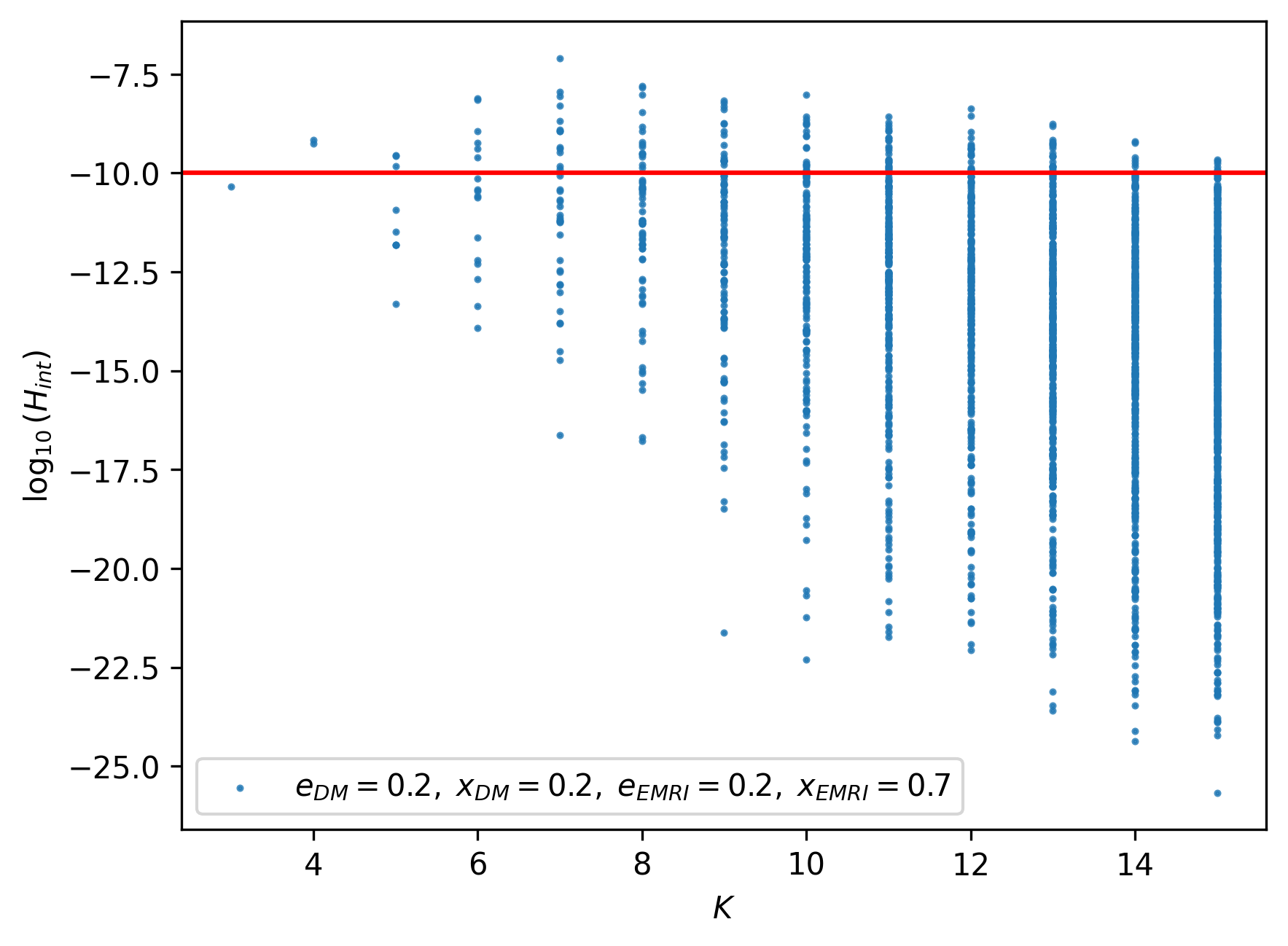}
  \caption{\(x_{\rm DM}=0.2\)}
  \label{fig:hint_m_eE0.2_xE0.7_eDM0.2_xDM0.2.png}
\end{subfigure}
\begin{subfigure}{0.48\textwidth}
  \centering
  \includegraphics[width=\linewidth]{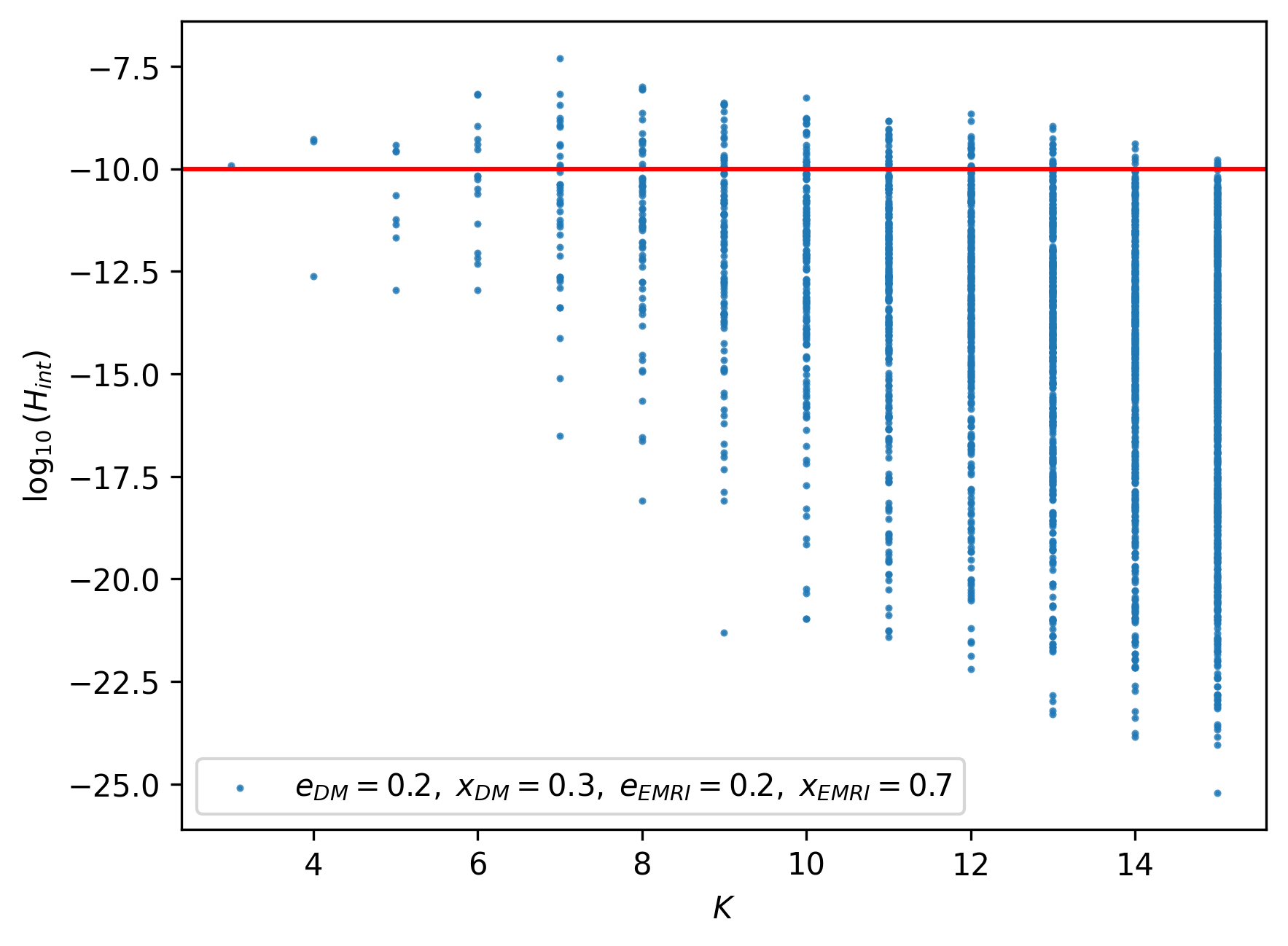}
  \caption{\(x_{\rm DM}=0.3\)}
  \label{fig:hint_m_eE0.2_xE0.7_eDM0.2_xDM0.3.png}
\end{subfigure}
  \caption{Decay of the interaction Hamiltonian \(H_{\rm int}\) as a function of \(\lvert K\rvert\) for \(x_{\rm DM}=0.1\), 0.2, and 0.3. The three curves—computed with all other parameters fixed as in Fig.\ref{fig:pDM_pEMRI.jpg} the progressively slower decay of \(H_{\rm int}\) at higher inclinations.}
  \label{fig:appendix1-3}
\end{figure}

\begin{figure}[htbp]
  \centering
\begin{subfigure}{0.48\textwidth}
  \centering
  \includegraphics[width=\linewidth]{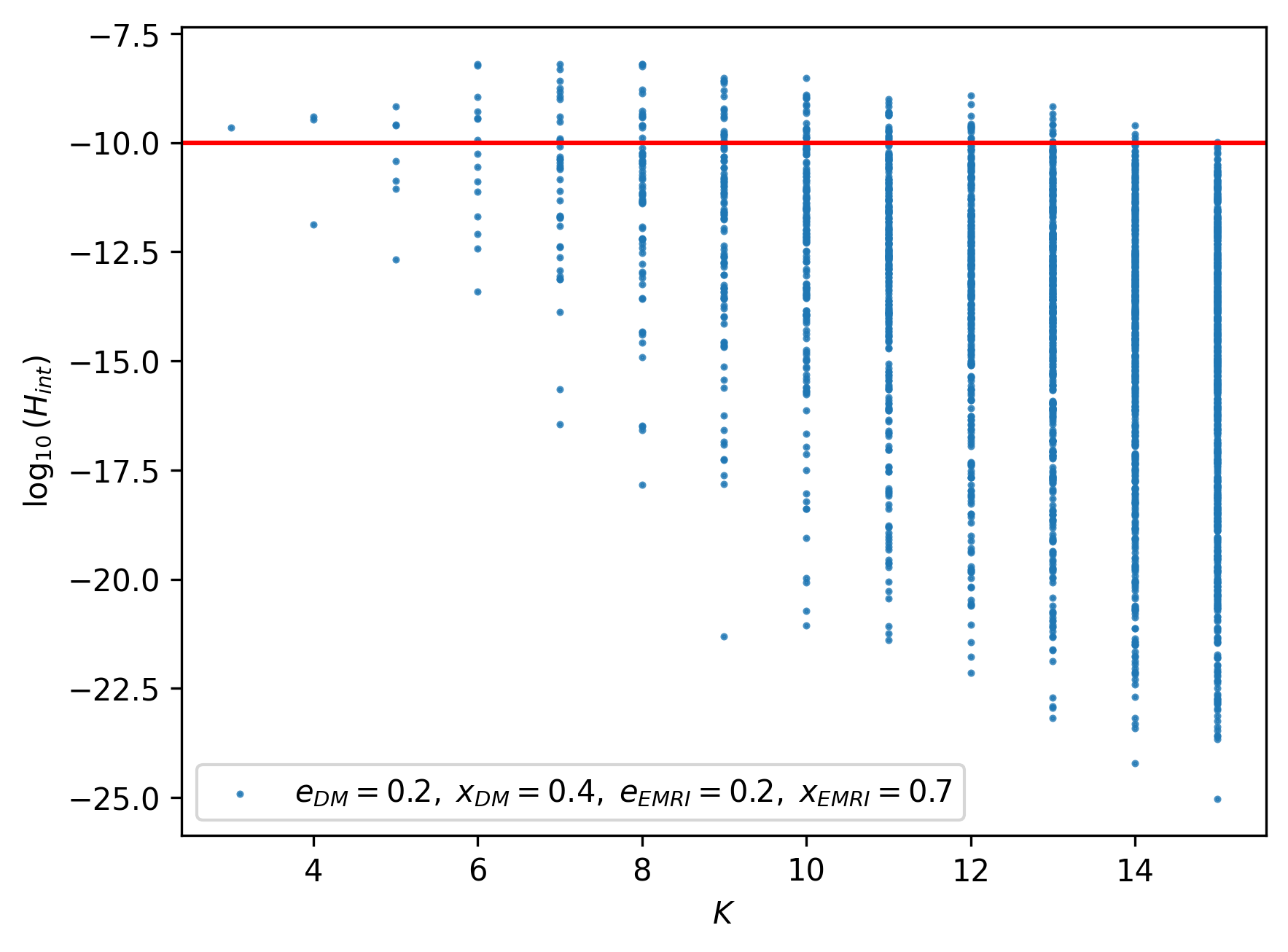}
  \caption{\(x_{\rm DM}=0.4\)}
  \label{fig:hint_m_eE0.2_xE0.7_eDM0.2_xDM0.4.png}
\end{subfigure}
\begin{subfigure}{0.48\textwidth}
  \centering
  \includegraphics[width=\linewidth]{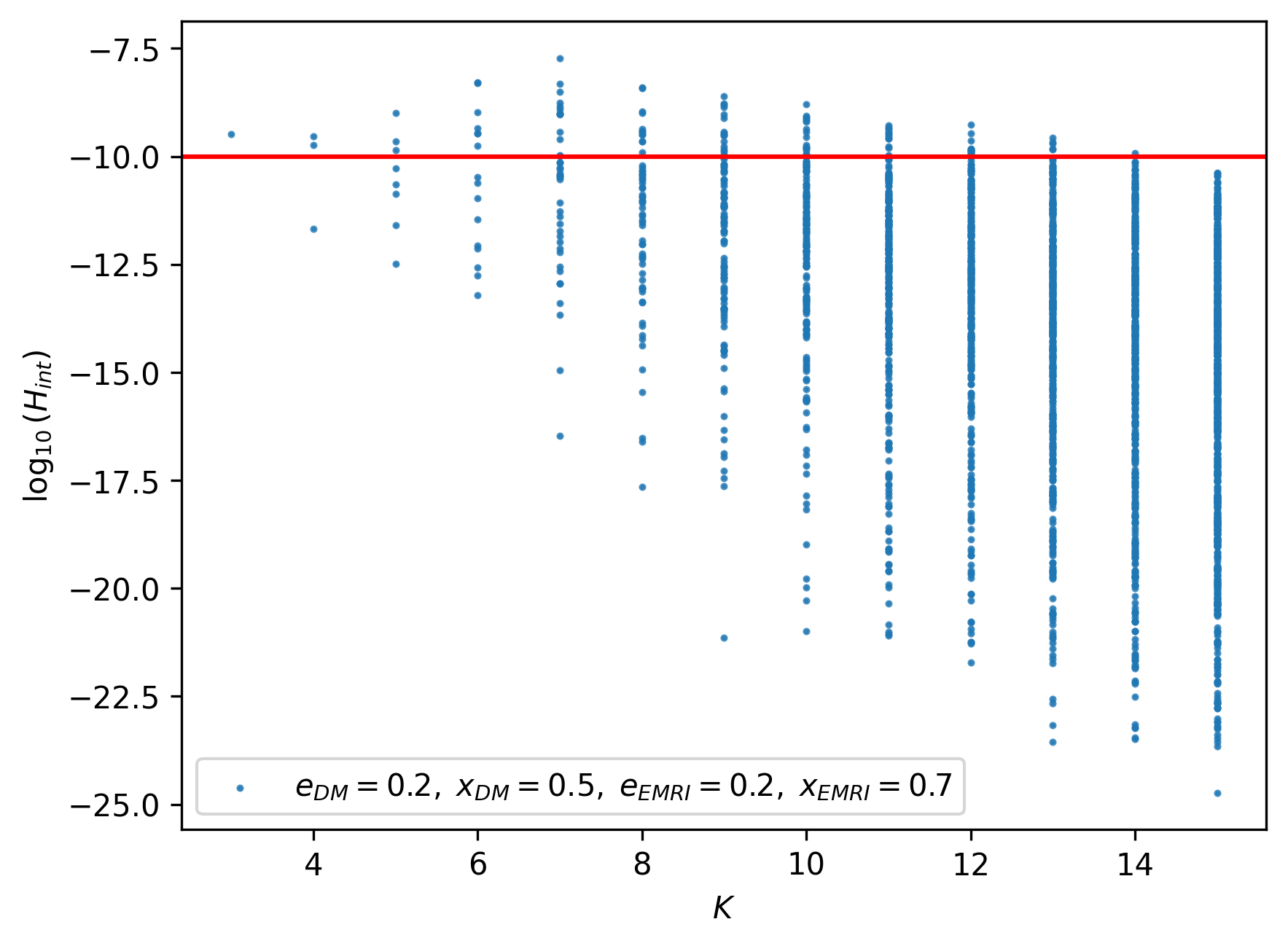}
  \caption{\(x_{\rm DM}=0.5\)}
  \label{fig:hint_m_eE0.2_xE0.7_eDM0.2_xDM0.5.png}
\end{subfigure}
\begin{subfigure}{0.48\textwidth}
  \centering
  \includegraphics[width=\linewidth]{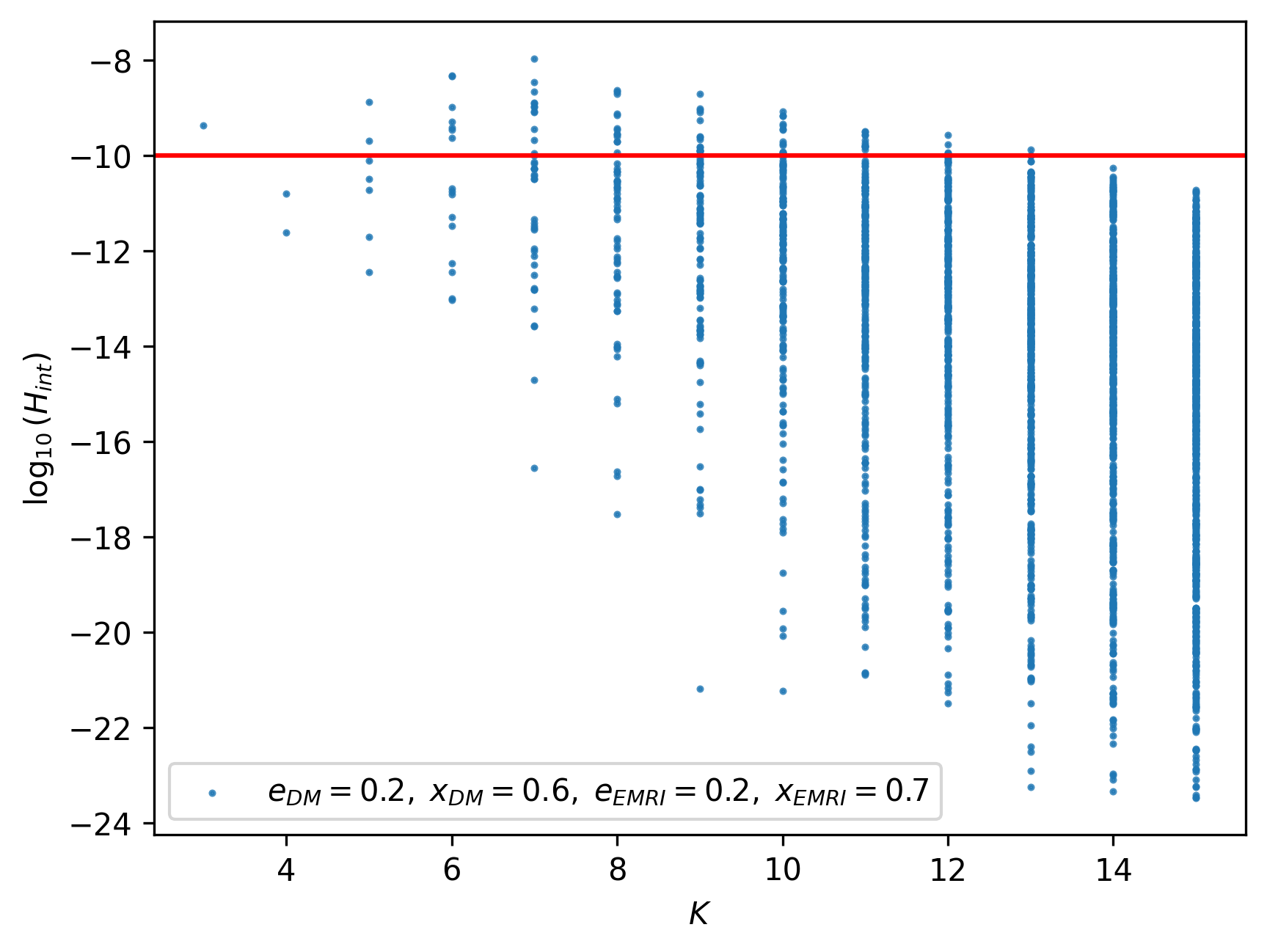}
  \caption{\(x_{\rm DM}=0.6\)}
  \label{fig:hint_m_eE0.2_xE0.7_eDM0.2_xDM0.6.png}
\end{subfigure}
  \caption{Same as in Fig.~\ref{fig:appendix1-3}, but restricted to the range \(x_{\rm DM}=0.4\)–0.6.}
  \label{fig:appendix4-6}
\end{figure}

We examined how the resonance amplitude \(H_{\rm int}\) varies with \( K\) for small values of \(x_{\rm DM}\) across a range of parameter sets in Figs.~\ref{fig:appendix1-3} and \ref{fig:appendix4-6}. The baseline parameters are those specified in Fig.~\ref{fig:pDM_pEMRI.jpg}.As the orbital inclination increases (i.e. as \(x\) decreases), the decay of the interaction Hamiltonian \(H_{\rm int}\) with respect to \( K\) becomes progressively poorer, rendering numerical evaluation more challenging. In this work, we therefore truncate the sum at \( K_{\max}=15\), for which convergence of the results has been verified.

\bibliographystyle{apsrev4-1}
\bibliography{refer}

\end{document}